\documentclass[journal,11pt,onecolumn,draftclsnofoot,]{IEEEtran}

\usepackage{graphicx}                % For including figures
\usepackage{amsmath}                 % For advanced math formatting
\usepackage{amsfonts}                % For additional math fonts
\usepackage{amssymb}                 % For more math symbols
\usepackage{cite}                    % For managing citations
\usepackage{hyperref}                % For hyperlinks in the document
\usepackage{tabularx}                % For flexible table formatting
\usepackage{multirow}                % For multi-row cells
\usepackage{booktabs}                % For professional table lines
\usepackage{float}                   % For [H] placement option
\usepackage{array}
\usepackage{xcolor} % Optional for cell shading
\usepackage{adjustbox} % For table adjustments
\usepackage{wrapfig} % For wrapping figures or tables

\usepackage{geometry}
\usepackage{amsmath}
\usepackage{subcaption}
\usepackage{booktabs}
\usepackage{makecell} % Allows line breaks in table cells
\usepackage{caption}     % Allows more control over captions
\usepackage[table]{xcolor}
\usepackage{colortbl}

\usepackage{soul}

\usepackage[normalem]{ulem}

\usepackage{lscape}

\usepackage{float}

\usepackage{tabularray}
\usepackage{algorithm}
\usepackage{algorithmic}
\usepackage[USenglish]{babel}
\usepackage[nodayofweek,level]{datetime}
\usepackage[utf8]{inputenc}
\usepackage{rotating}

\usepackage{pdflscape}
\usepackage{afterpage}
\usepackage{capt-of}% or use the larger `caption` package

\usepackage{lipsum}% dummy tex
\usepackage{cite}
\usepackage{csquotes}

\usepackage{tabu}
\def\tsc#1{\csdef{#1}{\textsc{\lowercase{#1}}\xspace}}
\tsc{WGM}
\tsc{QE}
\tsc{EP}
\tsc{PMS}
\tsc{BEC}
\tsc{DE}
\usepackage{framed}

\usepackage{flushend}

\usepackage{bm}

\usepackage{nomencl}
\usepackage{bbm}

\renewcommand\nomgroup[1]{%
    \item[\bfseries
     \ifstrequal{#1}{O}{\textbf{Operators}}{%
        \ifstrequal{#1}{S}{\textbf{Indices and sets}}{%
            \ifstrequal{#1}{V}{\textbf{Variables and parameters}}{%
            }}}
            ]}
            
            \makenomenclature

\makeatletter
\newsavebox\myboxA
\newsavebox\myboxB
\newlength\mylenA

\newcommand*\xoverline[2][0.75]{%
    \sbox{\myboxA}{$\m@th#2$}%
    \setbox\myboxB\null% Phantom box
    \ht\myboxB=\ht\myboxA%
    \dp\myboxB=\dp\myboxA%
    \wd\myboxB=#1\wd\myboxA% Scale phantom
    \sbox\myboxB{$\m@th\overline{\copy\myboxB}$}%  Overlined phantom
    \setlength\mylenA{\the\wd\myboxA}%   calc width diff
    \addtolength\mylenA{-\the\wd\myboxB}%
    \ifdim\wd\myboxB<\wd\myboxA%
      \rlap{\hskip 0.5\mylenA\usebox\myboxB}{\usebox\myboxA}%
    \else
        \hskip -0.5\mylenA\rlap{\usebox\myboxA}{\hskip 0.5\mylenA\usebox\myboxB}%
    \fi}
\makeatother
\makenomenclature

\usepackage{subcaption}
\usepackage{mleftright}
\usepackage{hyperref}

\usepackage{amsmath}
\usepackage{amssymb}
\usepackage{amsfonts}

\usepackage{mdframed}

\newmdenv[leftline=false,rightline=false,linewidth=1pt]{topbot2}

\usepackage{forest}
\usetikzlibrary{external}
\let\oldtikzexternalgetnextfilename\tikzexternalgetnextfilename \renewcommand{\tikzexternalgetnextfilename}[1]{\oldtikzexternalgetnextfilename{#1}\expandafter\tikzsetnextfilename\expandafter{#1}}

\usepackage{pgfplotstable}
\usepackage[outline]{contour}
\contourlength{1.2pt}

\pgfplotsset{compat=1.13} 

\usetikzlibrary{spy}
\usetikzlibrary{calc}
\usetikzlibrary{fadings}
\usetikzlibrary{patterns}
\usetikzlibrary{shadows}
\usetikzlibrary{mindmap}
\usetikzlibrary{backgrounds}
\usetikzlibrary{shapes.symbols}
\usetikzlibrary{shapes.multipart}
\usetikzlibrary{shapes.geometric}
\usetikzlibrary{automata,positioning}
\usetikzlibrary{decorations.fractals} 
\usetikzlibrary{decorations.markings}
\usetikzlibrary{decorations.pathreplacing}
\usetikzlibrary{decorations.pathmorphing}
 
\usepackage{bm}

\usepackage{nomencl}
\makenomenclature
\tikzset{edge from parent/.style={segment angle=10,draw}}

\tikzset{
 my rounded corners/.append style={rounded corners=2pt},
}

\def\BibTeX{{\rm B\kern-.05em{\sc i\kern-.025em b}\kern-.08em
 T\kern-.1667em\lower.7ex\hbox{E}\kern-.125emX}}

\renewcommand{\nomgroup}[1]{%
 \ifthenelse{\equal{#1}{O}}{\item[\textit{Operators}]}{%
 \ifthenelse{\equal{#1}{I}}{\item[\textit{Indices}]}{%
 \ifthenelse{\equal{#1}{A}}{\item[\textit{Acronyms}]}{%
 `\ifthenelse{\equal{#1}{V}}{\item[\textit{Variables and parameters}]}{}}}}}
\usepackage{scalerel}
\usetikzlibrary{svg.path}
\definecolor{orcidlogocol}{HTML}{A6CE39}
\tikzset{
 orcidlogo/.pic={
 \fill[orcidlogocol] svg{M256,128c0,70.7-57.3,128-128,128C57.3,256,0,198.7,0,128C0,57.3,57.3,0,128,0C198.7,0,256,57.3,256,128z};
 \fill[white] svg{M86.3,186.2H70.9V79.1h15.4v48.4V186.2z}
 svg{M108.9,79.1h41.6c39.6,0,57,28.3,57,53.6c0,27.5-21.5,53.6-56.8,53.6h-41.8V79.1z M124.3,172.4h24.5c34.9,0,42.9-26.5,42.9-39.7c0-21.5-13.7-39.7-43.7-39.7h-23.7V172.4z}
 svg{M88.7,56.8c0,5.5-4.5,10.1-10.1,10.1c-5.6,0-10.1-4.6-10.1-10.1c0-5.6,4.5-10.1,10.1-10.1C84.2,46.7,88.7,51.3,88.7,56.8z};
 }
}

\newcommand\orcidicon[1]{\href{https://orcid.org/#1}{\mbox{\scalerel*{ \begin{tikzpicture}[yscale=-1,transform shape]
 \pic{orcidlogo};
 \end{tikzpicture}
 }{|}}}}

\usepackage{booktabs}
\usepackage{threeparttable}
\usepackage{caption}
\usepackage{siunitx}

\begin{document}

% Title and authors
% \title{Uncertainty quantification and propagation 
% for residential, industrial and office building load profiles with increased adoption of electric vehicles and photovoltaic generation}

\title{\huge{Uncertainty quantification in load profiles with rising EV and PV adoption: the case of residential, industrial, and office buildings}}
% \title{Uncertainty quantification in load profiles with rising EV and PV adoption: the case of residential, industrial, and commercial consumers}

\author{
% ,~\IEEEmembership{Student~Member,~IEEE}~\orcidicon{0000-0001-7776-172X},
Aiko~Fias\textsuperscript{\textdagger},
~Md~Umar~Hashmi\textsuperscript{\textdagger},~\IEEEmembership{Senior~Member~IEEE}~\orcidicon{0000-0002-0193-6703},
and~Geert~Deconinck\textsuperscript{\textdagger},~\IEEEmembership{Senior~Member~IEEE}~\orcidicon{0000-0002-2225-3987}
\thanks{Corresponding author email: mdumar.hashmi@kuleuven.be}
\thanks{\textsuperscript{\textdagger} A.F., M.U.H., and G.D. are with KU Leuven \& EnergyVille, Belgium.}
% \thanks{This work is supported by the Flemish Government and Flanders Innovation \& Entrepreneurship (VLAIO) through the Flux50 project IMPROcap (HBC 2022.0733), {KU Leuven funded "FlexIQ" project (C2M/24/028)} and  partially funded as DigiRES under CETPartnership-2023 by the European Commission (Grant 101069750) and VLAIO (CETP-2023-00493).}}
}
\maketitle
% Abstract
\begin{abstract}
The integration of photovoltaic (PV) generation and electric vehicle (EV) charging introduces significant uncertainty in electricity consumption patterns, particularly at the distribution level. This paper presents a comparative study for selecting metrics for uncertainty quantification (UQ) for net load profiles of residential, industrial, and office buildings under increased DER penetration. A variety of statistical metrics is evaluated for their usefulness in quantifying uncertainty, including, but not limited to, standard deviation, entropy, ramps, and distance metrics. 
\textcolor{black}{The proposed metrics are classified into baseline-free, with baseline and error-based. These UQ metrics are evaluated for increased penetration of EV and PV. }
% Scenario generation is performed using real-world datasets, and stochastic modelling of EV charging is also considered. 
The results highlight suitable metrics to quantify uncertainty per consumer type and demonstrate how net load uncertainty is affected by EV and PV adoption. Additionally, it is observed that joint consideration of EV and PV can reduce overall uncertainty due to compensatory effects of EV charging and PV generation due to temporal alignment during the day. Uncertainty reduction is observed across all datasets and is most pronounced for the office building dataset.\\
\end{abstract}
% Keywords
\vspace{20pt}


\pagebreak

\tableofcontents

\pagebreak

% Main content of the paper
\section{Introduction}
% The decarbonization of power systems has led to increasing integration of DERs such as PV and EVs, particularly in low-voltage distribution networks. These technologies, while beneficial from an environmental standpoint, introduce variable and user-driven behaviors that challenge traditional grid planning and operation. As deterministic models fall short under such variability, Uncertainty quantification (UQ) becomes essential to characterize net load behavior.

% This paper investigates which UQ metrics are most effective for different types of consumers and how uncertainty evolves with increased PV and EV penetration. Unlike many existing studies, this work explicitly models EV and PV together in the net load, revealing potential interactions that influence uncertainty levels

% \subsection{Motivation}
% The paper aims to answer:
% \begin{itemize}
%     \item What uncertainty \textbf{quantification metrics} should be used for different kinds of electricity load profiles with increased penetration of EV and PV?
%     \item Assessing the \textbf{uncertainty propagation sensitivity} of load profiles towards increased EV and PV penetration?
%     \item Provide recommendations
% \end{itemize}

\IEEEPARstart{T}{he} energy sector is undergoing a rapid transformation driven by the urgent need to decarbonize. In this transition, distributed energy resources (DERs) such as photovoltaic (PV) systems and electric vehicles (EVs) play a crucial role \cite{PotentialDistributedEnergya}. Their use supports climate goals. However, these technologies also introduce new issues. Unlike traditional centralized power generation, DERs are unpredictable and controlled by users, making them inherently uncertain, introducing challenges for grid operators in terms of operation and planning \cite{ReviewOperationalChallenges}. 
\textcolor{black}{
Further, the high simultaneity factor\footnote{\textcolor{black}{Simultaneity factor of DERs measures the probability/extent to which multiple DERs (e.g. PV, EV, batteries etc.) produce or consume at the same time.}} of DERs showcased in \cite{fani2024impact} for EVs and in \cite{hashmi2023robust} for PV generation necessitates the impact assessment of large-scale integration of DERs.}

At the distribution level, high DER penetration alters traditional load and generation patterns. For instance, PV systems inject variable power based on solar irradiance conditions, while EVs can create sharp, localized demand surges during charging periods. As a result, traditional deterministic grid planning methods are no longer sufficient. Distribution networks (DNs) must be designed and operated under uncertainty, accounting for probabilistic variations in both load and generation.

To address this, uncertainty quantification (UQ) methods have gained attention \cite{deterministic_probabilistic}. These methods aim to characterize and represent the uncertainty inherent in stochastic variables, such as net load, in order to support system operation and planning under uncertainty. 
% This requires modelling the uncertainty. 
To model the uncertainty, it is crucial to understand how to quantify it. 
This motivates the present work, where we aim to identify UQ metrics for growing PV and EV penetration.
UQ methods are typically classified as aleatory and epistemic \cite{der2009aleatory}. In this work, only aleatory UQ is considered, as we do not assume additional parameter measurement other than consumer load profile time-series data for UQ.

\subsection{Literature Review}
UQ in load profiles is often achieved through statistical metrics such as mean and standard deviation. These metrics can be used for example to fit parametric distributions to historical data. For example, \cite{AnalysisVoltageRegulation} models load currents using the beta distribution. To fit this distribution to the historical data, only the mean, standard deviation, minimum and maximum are needed \cite{ProbabilisticModelResidential}. Another study using the beta distribution is \cite{UQ_LV}. 
\cite{CharacterizingProbabilityDensity} evaluates the Weibull and log-normal probability distributions for household electricity consumption. It used the mean and standard deviation of the data and the fitted distribution to evaluate variations between the two. In \cite{DataDrivenProbabilisticPower}, the uncertainty in electricity demand is represented by a Gaussian distribution. The Gaussian distribution is fully characterized by the mean and standard deviation. 

The research on uncertainty in load profiles is not solely focused on using statistical metrics to fit probability distributions. \cite{DailyLoadProfilesa}, for example, conducted a statistical analysis of load curves for residential, commercial, and industrial consumers. The study recommends using representative load curves, which are defined statistically by the mean and standard deviation. \cite{DistributionTransformerLoadinga} and \cite{ModelingLoadUncertainty} also used mean and standard deviation to represent uncertainty in daily load profiles.
Additionally, this last study uses entropy as a measure to \textcolor{black}{quantify} uncertainty. 
Authors in \cite{bandoria2025modeling} use 705 building load profiles to training for predicting standard deviation curves, as a que for uncertainty quantification of 2250 unseen curves from 30 buildings.
Table
\ref{tab:overview_literature} summarizes a few studies and highlights the features of the data used for UQ.
% to characterize the load profile uncertainty. 

\begin{table}[htbp]
    \caption{Used metrics in literature to quantify uncertainty}
    \vspace{-5pt}
    \centering
    \resizebox{0.7\columnwidth}{!}{
    \begin{tabular}{cccccccc}
    \toprule
    \textbf{Paper}  & \textbf{Mean} & \textbf{Std} & \textbf{Higher} & \textbf{Monthly} &  \textbf{Min/} & \textbf{Entropy} \\
    \textbf{ID}  &  & \textbf{Dev} & \textbf{moments} & \textbf{consumption} &  \textbf{Max} & \\
    \midrule    
    \cite{AnalysisVoltageRegulation} & \checkmark & \checkmark & &  & \checkmark &   \\
    \cite{UQ_LV} & \checkmark & \checkmark & &  & \checkmark &   \\
    \cite{CharacterizingProbabilityDensity} & \checkmark & \checkmark & &  &  &   \\
    \cite{DataDrivenProbabilisticPower} & \checkmark & \checkmark & &  &  &   \\
    \cite{DailyLoadProfilesa} & \checkmark & \checkmark & &  \checkmark &  &   \\
    \cite{ModelingLoadUncertainty} & \checkmark & \checkmark & &  &  & \checkmark \\
    \cite{StudyDistributionCharacteristics} &  &  & &  &  & \checkmark \\
    \cite{DistributionTransformerLoadinga} &  \checkmark & \checkmark & &  \checkmark &  & \\
    \cite{UncertaintyAwareComputationalTools} &  \checkmark & \checkmark & &   &  & \\
    \cite{QuantifyingLoadUncertaintya} &  \checkmark &  & &   & \checkmark & \\
    \cite{skew_kurt} &  \checkmark & \checkmark & \checkmark &   &  & \\
    \cite{UncertaintyModelingDistributed}  &  \checkmark & \checkmark &  &   &  & \\
    \bottomrule
    \end{tabular}}
    \label{tab:overview_literature}
\end{table}

A lot of research considering DERs, takes either EV or PV  into account in their analysis. Studies such as \cite{DataDrivenProbabilisticPower}, \cite{FastspecializedPointEstimate}, and \cite{skew_kurt} 
analyzed the impact of PV uncertainties on the network.
Other studies like \cite{UncertaintyAwareComputationalTools},  \cite{ImpactElectricVehicles}, and \cite{ProbabilityEvaluationExcess}, 
have addressed the uncertainties associated with EV in their analysis.
However, many studies start to focus on how these uncertainties affect distribution grids when both PV and EV are present simultaneously. For example, \cite{thesis_Milan} looked at both PV and EV uncertainties to analyze their impact on the low voltage network. The study found that the demand by EV is compensated by the power injections of PV,
% . Their uncertainties are mitigating each other, which 
\textcolor{black}{in effect causes a lower burden on the DN.}
% Another study \cite{ProbabilisticLoadFlowb} also used probabilistic load flow techniques to analyze the effect of both EV and PV uncertainties. Lastly, \cite{RiskbasedSimulationMultiobjective} investigated how to find the best size and placement of distributed energy sources, taking into account the uncertainty from both PV and EV systems.

Despite the variety of approaches in existing literature, several important gaps remain. Most studies rely on basic statistical metrics, primarily the mean and standard deviation, to quantify the uncertainty, which may not fully capture the complexity of load variability under high DER penetration. Furthermore, still a lot of works focus on either EV or PV separately, neglecting the combined and potentially interacting effects of \textcolor{black}{multiple} DERs on net load uncertainty. 

\subsection{Contributions}
This paper addresses the above research gaps through a comprehensive uncertainty quantification study focused on net load profiles impacted by both PV and EV penetration. The contributions of this work are the following:
\begin{itemize}
    \item 
    % Identifying which uncertainty \textbf{quantification metrics} should be used for different kinds of electricity load profiles with increased penetration of EV and PV. 
    A variety of \textbf{uncertainty quantification metrics}, classified as baseline-free, baseline-dependent and error-based metrics, are evaluated based on their usefulness in quantifying uncertainty in behind-the-meter load profiles with increased EV or PV penetration for residential consumers, industrial consumers and office building load. 
    This leads to recommendations on suitable metrics for each consumer type.
    % This results in recommendations on which metrics to use for which type of consumer.
    
    \item 
    The \textbf{uncertainty sensitivity} of the different types of load profiles towards per-unit increase in EV and/or PV penetration is analyzed. The results indicate which DER contributes the most to specific aspects of uncertainty.

    \item The impact on uncertainty when considering \textbf{EV and PV together} in the net load rather than adding there effects separately is evaluated.
    The results show that EV and PV interact with each other, reducing the uncertainty in the load profile. 
    The reductions are the highest when EV consumption overlaps with the PV generation. 
    % \textcolor{red}{To add qualitative results on how much STD and peaks reduced!!}
\end{itemize}

Through these contributions, we aim to provide actionable insights for \textcolor{black}{distribution system operators (DSOs), aggregators, flexibility service providers (FSP) and other stakeholders} to develop robust uncertainty-aware planning and operational strategies for DER-integrated power distribution systems.

\pagebreak

\section{Uncertainty quantification metrics} \label{sec:metrics}
\begin{landscape}
\begin{figure*}[htbp]
    \centering
    \includegraphics[width=1.1\columnwidth]{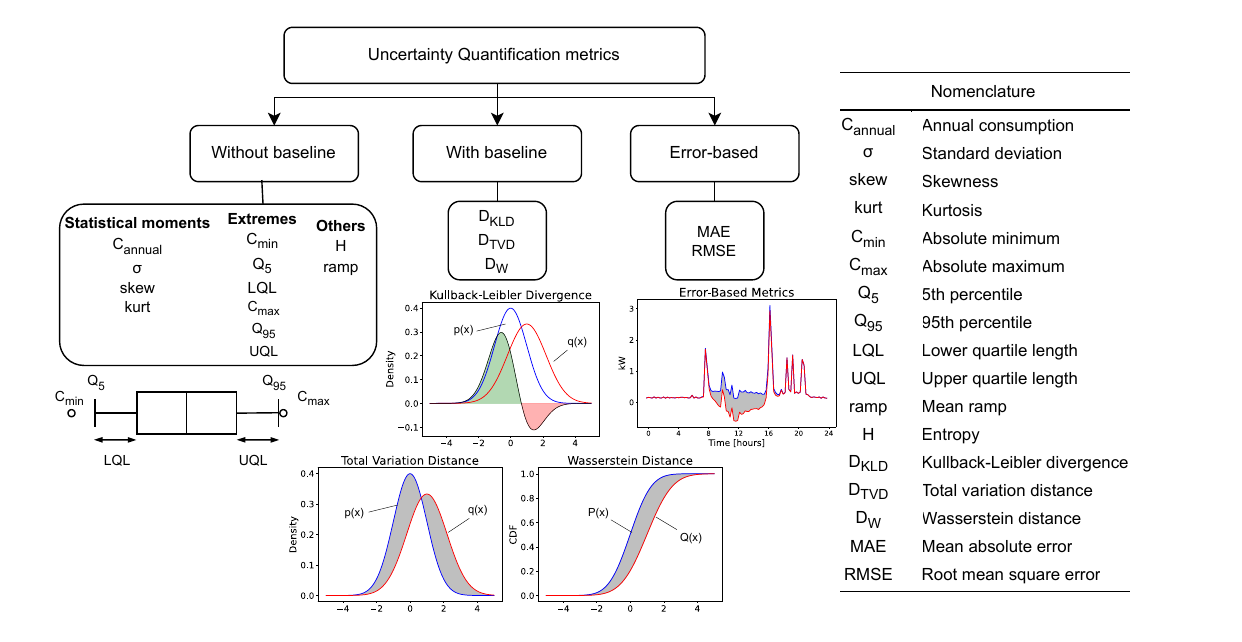}
    \vspace{-4pt}
    \caption{Overview of UQ metrics based on three classifications: (i) without baseline, (ii) with baseline,  (iii) error-based metrics.}
    \label{fig:overview_metrics}
\end{figure*}
\end{landscape}

Various metrics to quantify uncertainty of net load profiles are evaluated. The metrics are divided into three main categories. 
% Fig. \ref{fig:overview_metrics} provides an overview of the metrics and their categories.
\textcolor{black}{
Fig. \ref{fig:overview_metrics} provides the nomenclature and classification of the UQ metrics. Note that metrics without a baseline rely only on the net load time series with PV/EV/both. However, metrics with a baseline and error-based metrics rely not only on the net load time series with PV/EV/both, but also need the net load time series without DERs. For the latter case, sub-metering of the DERs is necessary.}
The following sections discuss each category.

\subsection{Metrics Without a Baseline}
These metrics evaluate uncertainty based solely on the profile itself. It includes the annual consumption, average daily standard deviation, average daily skewness and kurtosis, annual mean ramp, average daily Shannon entropy, yearly minimum, maximum, 5th and 95th percentile and average daily upper and lower quartile length.
The equations are given respectively in \eqref{eq:load}-\eqref{eq:right tail}. In these equations $x_{d,t}$ is the power consumed in kW at time step $t \in \{1,.., T\}$ on day $d \in \{1,..,D\}$, where $D$ is the number of days, $T$ the amount of time steps in a day and $N = D \cdot T$ is the total number of data points in the yearly profile. Furthermore, $p_d$ is the discrete probability density function (PDF) of day $d$, consisting of $B_d$ bins. Finally, $Q_i$ represents the \textit{i}-th percentile.
\begin{equation}
    C_{\text{annual}} = \sum_{d=1}^{D}\sum_{t=1}^{T} x_{d,t} \cdot \Delta t \label{eq:load} 
\end{equation}
\begin{equation}
    \sigma =\frac{1}{D} \sum_{d=1}^{D}  \sqrt{\frac{1}{T} \sum_{t=1}^{T} (x_{d,t}-\overline{x_d})^2} \label{eq:std day} 
\end{equation}
\begin{equation}
    skew = \frac{1}{D} \sum_{d=1}^{D} \frac{1}{T} \sum_{t=1}^{T} \frac{(x_{d,t}-\overline{x_d})^3}{\sigma_d^3} \label{eq:skew day}
\end{equation}
\begin{equation}
    kurt = \frac{1}{D} \sum_{d=1}^{D} \frac{1}{T} \sum_{t=1}^{T} \frac{(x_{d,t}-\overline{x_d})^4}{\sigma_d^4}-3 \label{eq:kurt day} 
\end{equation}
\begin{equation}
    ramp = \frac{1}{N-1} \sum_{t=2}^{N} |x_{t} - x_{t-1}| \label{eq:ramp} 
\end{equation}
\begin{equation}
    H = - \frac{1}{D} \sum_{d=1}^{D} \sum_{i=0}^{B_d} p_d(x_i) \log p_d(x_i) \label{eq:shannon day}
\end{equation}
\begin{equation}
    C_{\text{min}} = \min \limits_{d  \in \{1,.. D\}} \min \limits_{t  \in \{1,.. N\}} x_{d,t} \label{eq:min}
\end{equation}
\begin{equation}
    C_{\text{max}} = \max \limits_{d  \in \{1,.. D\}} \max \limits_{t  \in \{1,.. N\}} x_{d,t} \label{eq:max} 
\end{equation}
\begin{equation}
    LQL = \frac{1}{D} \sum_{d=1}^{D}  Q_{25}^d - Q_0^d  \label{eq:left tail} 
\end{equation}
\begin{equation}
    UQL = \frac{1}{D} \sum_{d=1}^{D}  Q_{100}^d - Q_{75}^d  \label{eq:right tail}
\end{equation}

\textcolor{black}{
$\Delta t$ denotes sampling time, $\overline{x_d}$ is the mean load for day $d$, H denotes Shannon entropy, LQL and UQL refer to lower and upper quartile lengths.}

\textcolor{black}{
The metrics without a baseline can be further classified into (i) statistical moments based (includes $C_{\text{annual}}, \sigma, skew, kurt$), (ii) extremes and quartiles (includes $C_{\min}, C_{\max}, LQL, Q_i$) and (iii) other metrics such as H and $ramp$, also shown in Fig. \ref{fig:overview_metrics}.
}

\subsection{Metrics with a Baseline}
These are metrics that are calculated relative to a defined baseline. In this context, the baseline refers to the base load profile without the presence of DERs. 
The metrics considered are the average daily Kullback-Leibler divergence (KLD), total variation distance (TVD) and Wasserstein distance between the PDF of the base load and the PDF of the net load (with DER penetration). The used formulas are given in \eqref{eq:KLD} - \eqref{eq:wass}. $p_d$ is the discrete PDF of the base load prior to adding DERs, and $q_d$ represents the discrete daily PDF of the net load that includes DERs. $P_d$ and $Q_d$ represent the daily cumulative distribution functions (CDF) of base and net load, respectively.
\begin{equation}
        D_{\text{KLD}} = \frac{1}{D} \sum_{d=1}^{D} \sum_{i=0}^{B_d} p_d(x_i) \log \frac{p_d(x_i)}{q_d(x_i)}
    \label{eq:KLD} 
\end{equation}
\begin{equation}
    D_{\text{TVD}} = \frac{1}{D} \sum_{d=1}^{D} \frac{1}{2} \sum_{i=0}^{B_d} |p_d(x_i) - q_d(x_i)|
    \label{eq:TVD} 
\end{equation}
\begin{equation}
    D_{\text{W}} = \frac{1}{D} \sum_{d=1}^{D} \sum_{i=0}^{B_d} |P_d(x_i) - Q_d(x_i)| \cdot \Delta x
    \label{eq:wass}
\end{equation}

\subsection{Error-based metrics}
Error-based metrics quantify the difference between two time series at each point in time. These metrics compare the actual values of the net load profile with those of the baseline (base load without DERs). The metrics considered are the mean absolute error and the root mean square error given by \eqref{eq:mae} and \eqref{eq:rmse}.
\begin{equation}
    \text{MAE} = \frac{1}{N} \sum_{t=1}^{N} \left| x_t - y_t \right| \label{eq:mae} 
\end{equation}
\begin{equation}
    \text{RMSE} = \sqrt{\frac{1}{N} \sum_{t=1}^{N} (x_t - y_t)^2} \label{eq:rmse} 
\end{equation}

\pagebreak

\section{Assessing UQ metrics}
\begin{landscape}
\begin{figure*}[htbp]
    \centering
    \includegraphics[width=1.1\columnwidth]{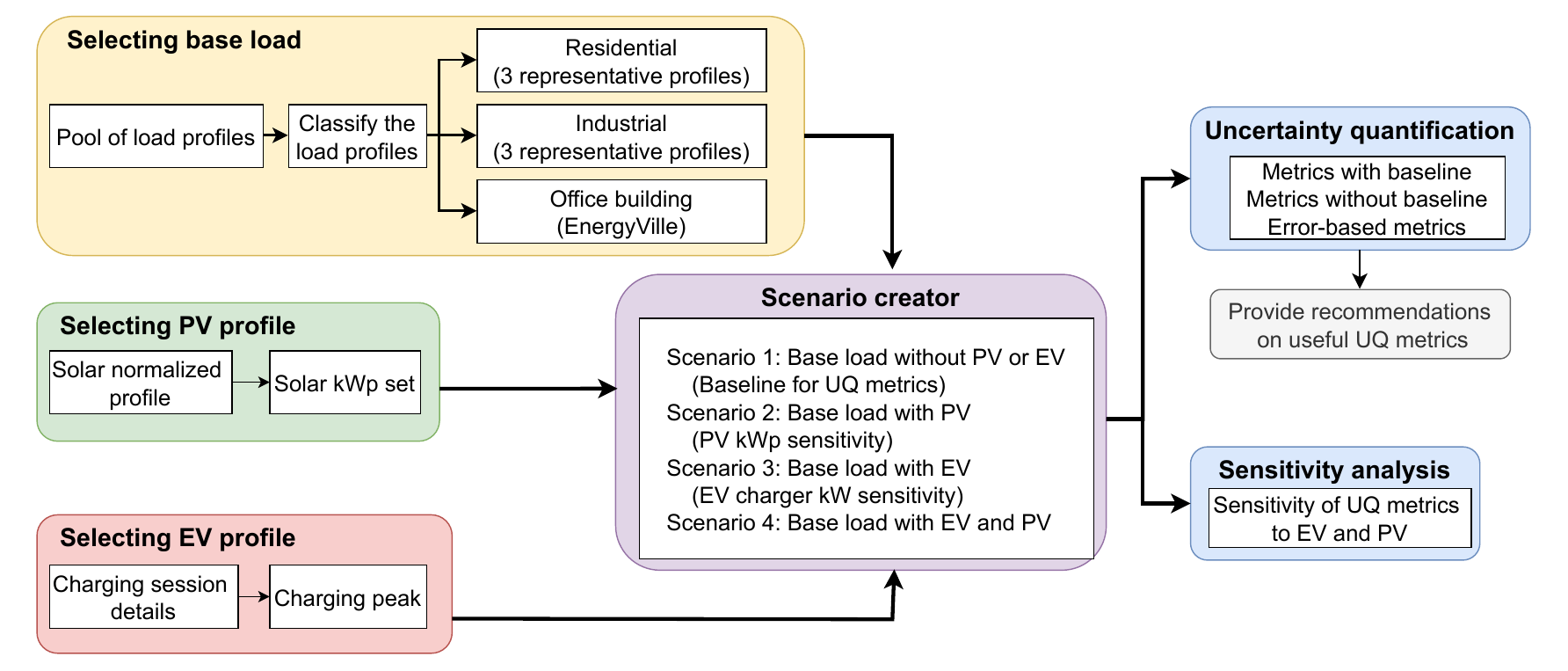}
    \vspace{-6pt}
    \caption{Proposed UQ framework with per unit change in EV and/or PV penetration for different load types}
    \label{fig:framework}
\end{figure*}
\end{landscape}

This section presents the 
assessment of UQ metrics detailed in Section \ref{sec:metrics}
for residential, industrial and office building load profiles with growing EV and PV levels. Fig. \ref{fig:framework} shows the flow of events.

% The UQ framework uses consumer datasets of three types: residential, industrial, and office buildings. 
For each load consumption dataset, DER integration scenarios are constructed by combining base load profiles with increasing levels of EV charging and PV generation. The resulting scenarios are then evaluated using the uncertainty quantification metrics, detailed in section \ref{sec:metrics}. Base load profiles are obtained from datasets representing historical electricity consumption without DERs. 
\textcolor{black}{
The UQ assessment relies on the source of data. Ideally, all load profile datasets (residential, industrial, and office building) and DER profiles should be from a single geographical site. However, it was not possible due to the lack of openly available datasets. The data sources are sensibly chosen from Belgium, 
% (residential consumer and office building load profile, solar PV generation), 
Germany 
% (industrial load profiles) 
and the Netherlands, 
% (EV profiles), 
see Tab. \ref{tab:datasource}. All these locations are in geographical proximity in central Europe and therefore can be assumed to have similar weather conditions and electricity consumption patterns.}

% Next, the subsections are dedicated to each consumer load profile types, i.e., residential, industrial and office building 
% % to address the selection of base 
% load profiles.
Charging profiles are generated using a stochastic model based on ElaadNL data, which captures statistical distributions of start times, session durations, and energy demands \cite{datasource}. Profiles are created daily and aggregated to form yearly profiles. PV generation data, obtained from Elia \cite{elia}, is normalized by installed capacity and assumed to be uniformly applicable across consumers. The normalized profile allows scaling for different levels of PV integration. A net load scenario combines base load, PV generation, and EV charging and is mathematically given by: $$\text{Net Load} = \text{Base Load} + \text{EV Charging} - \text{PV Generation}.$$
To evaluate the metrics, the base load profile and the charging behavior are kept fixed while the PV and EV levels are systematically increased. %This approach allows for isolating the effect of PV and EV on a specific consumer.
% For each net load scenario, the metrics are calculated.

\begin{table}
\centering
\caption{\textcolor{black}{Datasets used for UQ in this paper}}
\begin{tabular}{l|lllll}
\textbf{Data type }               & \textbf{Country}     & \textbf{Source}  & \textbf{Year}  & \textbf{Ref.} & \textbf{Open source}  \\ \hline
Solar PV      & Belgium     & Elia                & 2022            &     \cite{elia,datasource}      & Yes          \\
EV profile               & Netherlands & ElaadNL             & 2019            &    \cite{ElaadData,datasource}       & No longer    \\
Residential   & Belgium     & Fluvius             & 2022            &       \cite{Fluvius,datasource}    & Yes          \\
Industrial    & Germany     & --                  & 2016            &        \cite{data_industrial}   & Yes          \\
Office load          & Belgium     & EnergyVille         & 2024            &    \cite{energyvilleDcPortal}       & No          
\end{tabular}
\label{tab:datasource}
\end{table}

% \vspace{-14pt}

\subsection{Residential load}
The residential load profiles are based on a 2022 dataset from the Flemish DSO Fluvius \cite{Fluvius}. %containing 300 residential electricity consumers without EV or PV installations, sampled every 15 minutes during one year 
To analyze the impact of load profile uncertainty, three representative consumers, a small, medium, and large, are selected based on annual consumption. Their respective annual consumptions are approximately 1000 kWh, 2500 kWh, and 6500 kWh. These three sampled residential load profiles are incrementally penetrated with PV and EV, increasing linearly from 0 kW to a maximum of 7 kW. For each level of DER penetration, all \textcolor{black}{UQ} metrics are calculated. Table \ref{tab:metrics_residential} presents a summary of these metrics, showing values for three key scenarios: the base case with no DER penetration, the case with maximum PV penetration (7 kW), and the case with maximum EV penetration (7 kW). S, M and L indicate the small, medium and large consumer, respectively.
\begin{table}[!ht]
    \caption{Result of metrics for residential consumers}
    \label{tab:metrics_residential}
    \centering
    \begin{threeparttable}
    \resizebox{0.85\columnwidth}{!}{
    % \makebox[0.7\columnwidth][c]{
    \begin{tabular}{l|ccc|ccc|ccc}
    \toprule
    & \multicolumn{3}{c|}{\textbf{Base}} & \multicolumn{3}{c|}{\textbf{With PV (7 kW)}} & \multicolumn{3}{c}{\textbf{With EV (7 kW)}} \\
    \textbf{Metric} & \textbf{S} & \textbf{M} & \textbf{L} & \textbf{S} & \textbf{M} & \textbf{L} & \textbf{S} & \textbf{M} & \textbf{L} \\
    \midrule
    $C_{\text{annual}}$ & 1.00 & 2.50 & 6.48 & -8.86 & -7.36 & -3.38 & 5.56 & 7.06 & 11.04 \\
    ${\sigma}$ & 0.14 & 0.30 & 0.55 & 1.42 & 1.42 & 1.34 & 1.59 & 1.63 & 1.72 \\
    ${skew}$ & 5.22 & 3.86 & 1.95 & -0.55 & -0.34 & -0.14 & 3.41 & 3.19 & 2.67 \\
    ${kurt}$ & 35.95 & 19.38 & 5.95 & 1.10 & 1.42 & 1.09 & 11.25 & 9.88 & 7.18 \\
    ${ramp}$ & 0.06 & 0.10 & 0.18 & 0.14 & 0.17 & 0.23 & 0.18 & 0.23 & 0.30 \\
    ${H}$ & 2.42 & 2.61 & 2.31 & 4.61 & 4.29 & 3.29 & 2.82 & 2.95 & 2.64 \\
    $C_{\text{min}}$ & 0.04 & 0.07 & 0 & -6.94 & -6.87 & -6.79 & 0.04 & 0.07 & 0 \\
    $Q_{5}$ & 0.05 & 0.11 & 0.19 & -5.20 & -4.96 & -4.30 & 0.05 & 0.11 & 0.19 \\
    ${LQL}$ & 0.01 & 0.03 & 0.12 & 1.73 & 1.83 & 1.77 & 0.01 & 0.04 & 0.14 \\
    $C_{\text{max}}$ & 3.04 & 5.47 & 5.98 & 3.04 & 4.79 & 4.99 & 10.04 & 10.78 & 10.98 \\
    $Q_{95}$ & 0.24 & 0.81 & 1.74 & 0.18 & 0.34 & 1.18 & 7.05 & 7.14 & 7.28 \\
    ${UQL}$ & 1.05 & 1.67 & 2.04 & 0.90 & 1.09 & 1.72 & 6.17 & 6.34 & 6.29 \\
    ${D_{\text{KLD}}}$ & 0 & 0 & 0 & 10.40 & 31.89 & 20.47 & 1.93 & 2.54 & 1.28 \\
    ${D_{\text{TVD}}}$ & 0 & 0 & 0 & 0.47 & 0.46 & 0.41 & 0.10 & 0.10 & 0.10 \\
    ${D_{\text{W}}}$ & 0 & 0 & 0 & 1.13 & 1.13 & 1.13 & 0.52 & 0.52 & 0.52 \\
    MAE & 0 & 0 & 0 & 1.13 & 1.13 & 1.13 & 0.52 & 0.52 & 0.52 \\
    RMSE & 0 & 0 & 0 & 2.10 & 2.10 & 2.10 & 1.82 & 1.82 & 1.82 \\
    \bottomrule
    \end{tabular}}
    \begin{tablenotes}[flushleft]
    \scriptsize
    \item \textit{Units:} $C_{\text{annual}}$ in MWh; $H$, $D_{\text{KLD}}$ in bits; $D_{\text{TVD}}$, $skew$, $kurt$ are unitless; others in kW
    \end{tablenotes}
    \end{threeparttable}
\end{table}

\begin{figure}[htbp]
    \centering
    \begin{subfigure}[b]{0.65\columnwidth}
        \centering
        \includegraphics[width=\columnwidth]{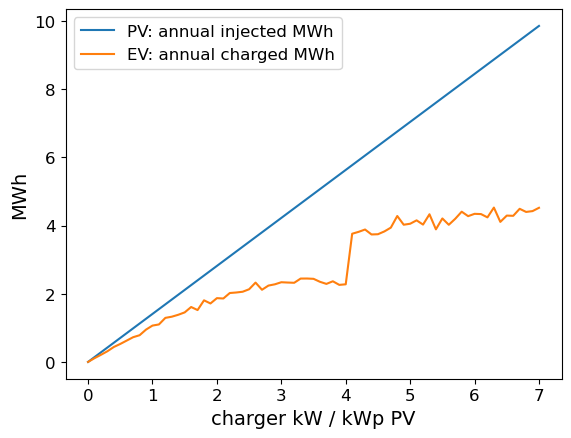}
        \caption{EV only and PV only}
        % \caption{Annual energy injected by PV and charged by EV}
        % \caption{Energy consumed by PV and EV}
        \label{fig:consumption pv and ev only}
    \end{subfigure}
    \begin{subfigure}[b]{0.71\columnwidth}
        \centering
        \includegraphics[width=\columnwidth]{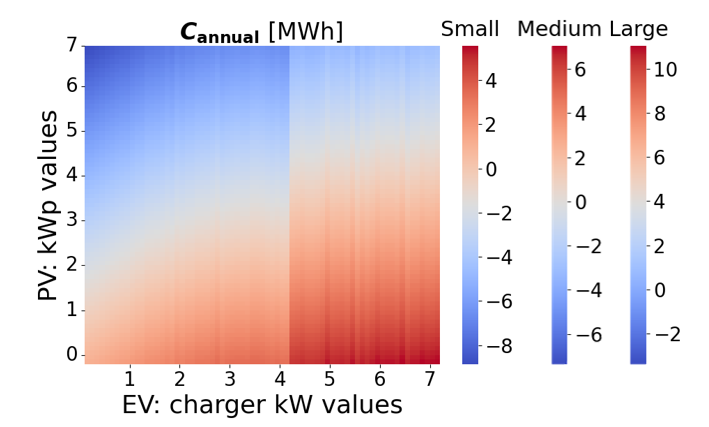}
        \vspace{0.5em}  % Adjust this value to move the caption up or down
        \caption{Net load}
        \label{fig:net cons}
    \end{subfigure}
    \caption{\small{\textcolor{black}{Annual energy injected by PV only, charged by EV only, and impact on net load}}}
    % \caption{Annual consumption of EV, PV and net load for different levels of PV and EV penetration}
    \label{fig: annual cons}
\end{figure}

The annual consumption ($C_{\text{annual}}$) increases linearly with the charged energy and decreases linearly with the installed capacity of PV. This results in large changes for all consumers since EV consumption and PV generation are relatively large compared to the small base load consumptions. The mean standard deviation (${\sigma}$) shows a large relative increase with EV and PV compared to the base load for the three consumers. It is increased proportionally with both EV and PV penetration since both DERs introduce large deviations from the mean daily consumption. Also, the mean ramp shows large relative increases compared to the base load since both EV and PV add fluctuations that are large relative to the inherent fluctuations in the profile. %PV due to the its variable nature and EV due to its large charging spikes.
These metrics are thus all useful to quantify uncertainty in the case of both increased EV and PV penetration.
\textcolor{black}{Fig. \ref{fig:consumption pv and ev only} shows the change in annual consumption for the PV generation and EV charging profiles for different levels of EV charger kW and PV installed kWp. The yearly injected energy due to PV generation increases linearly with installed kWp of PV. The yearly consumed energy due to EV has small fluctuations due to variations in charging profiles. 
Fig. \ref{fig:consumption pv and ev only} also shows that PV impacts $C_{\text{annual}}$ more compared to EV due to the linear scaling of PV generation with respect to PV size in kWp. However, the annual charged energy is not just a function of the charger size but is also limited by the EV battery size.
Fig. \ref{fig:net cons} is a heatmap showing the annual consumption of the net load for different levels of EV and PV penetration.
In this heatmap, the \textit{x}-axis shows the increasing charger kW values for EV, and the \textit{y}-axis shows the increasing kWp values for PV, both ranging from 0 to 7 kW. 
% The annual consumption of the net load is denoted as
% % the annual consumption of the base load added with the annual consumption due to EV and subtracted with the annual injected energy due to PV: 
% $C_{\text{annual}} = C_{base} + C_{EV} - C_{PV}$. 
All the consumers have the same trends when EV and PV are added, but at different scales. The heatmap in Fig. \ref{fig:net cons} shows the scales for the S, M and L consumers.}

% \vspace{-15pt}

\begin{figure}[htbp]
    \centering
    \begin{subfigure}{0.49\columnwidth}
        \centering
    \includegraphics[width=0.96\columnwidth]{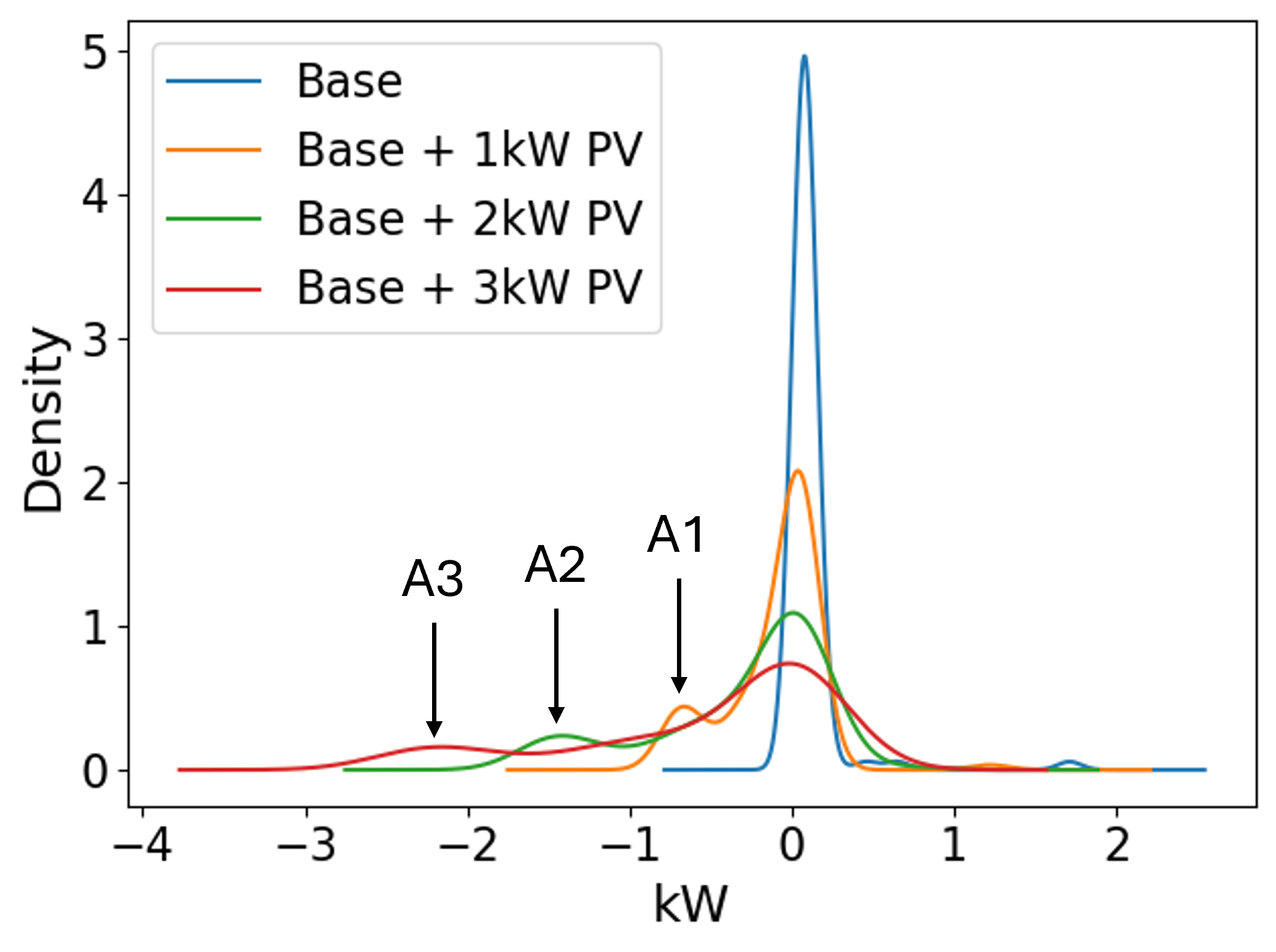}
    \end{subfigure}
    \hfill
    \begin{subfigure}{0.48\columnwidth}
        \centering
        \includegraphics[width=0.96\columnwidth]{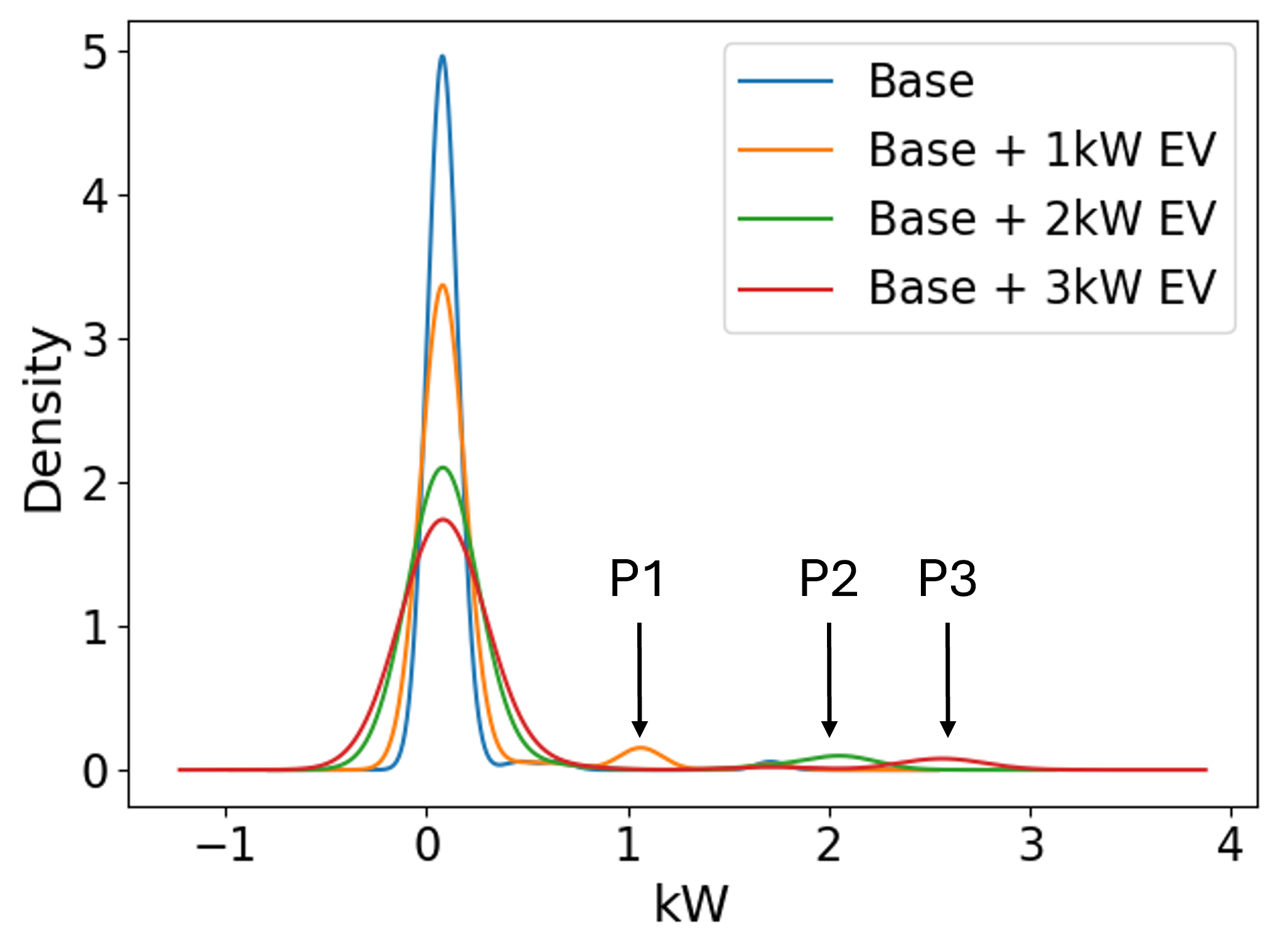}
    \end{subfigure}
    \caption{\small{Comparison of PDFs for different levels of PV and EV penetration for a small consumer}}
    \label{fig:PDF_pv_ev}
\end{figure}

The skewness and kurtosis experience a large decrease when either EV or PV is added. This happens because the addition of EV or PV introduces a second mode in the PDF. Fig. \ref{fig:PDF_pv_ev} shows this effect. The mode introduced by PV is indicated by A1, A2, and A3 for increasing PV levels. This mode corresponds to lower, negative consumption during peak solar generation. %The existing mode corresponds to higher consumption values during non-solar hours. 
The mode introduced by EV is indicated by P1, P2, and P3 for increasing EV levels. This second mode represents the small part of the profile with increased consumption that occurs when charging takes place. %When the profile contains both EV and PV, the same modes as in the scenarios with only PV are present. PV decreases the increased consumption due to EV, keeping it in the same range as the consumption of base load, which is why EV does not introduce an extra mode.% Due to the reduction in skewness and kurtosis, it might incorrectly appear that the distribution becomes more similar to a normal distribution, whereas the opposite is actually true. 
% Skewness and kurtosis are defined relative to a normal distribution, but i
In the case of a bimodal PDF, a distribution with near-zero skewness and kurtosis does not imply normality, and thus these metrics lose interpretability. Therefore, in the presence of bimodality, skewness and kurtosis are considered less meaningful indicators of the shape of the distribution. 

\begin{figure}[htbp]
    \centering
    \resizebox{\columnwidth}{!}{
    \begin{subfigure}{0.49\columnwidth}
        \centering
        \includegraphics[width=0.96\columnwidth]{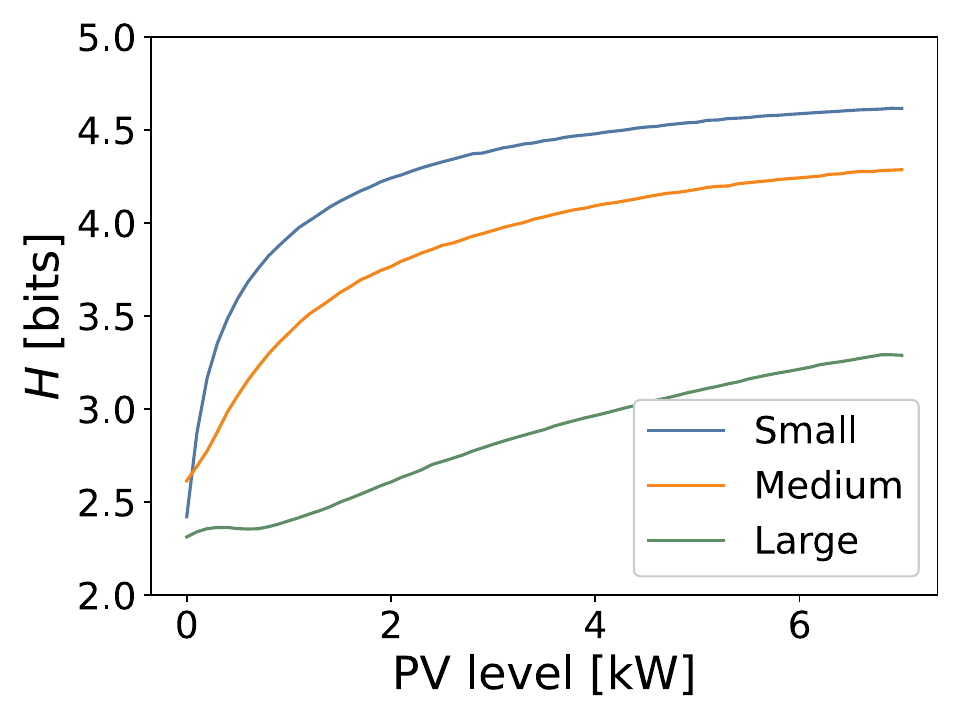}
        % \caption{Mean Shannon entropy for different levels of PV}
        % \label{fig:shannon_pv}
    \end{subfigure}
    \begin{subfigure}{0.48\columnwidth}
        \centering
        \includegraphics[width=0.96\columnwidth]{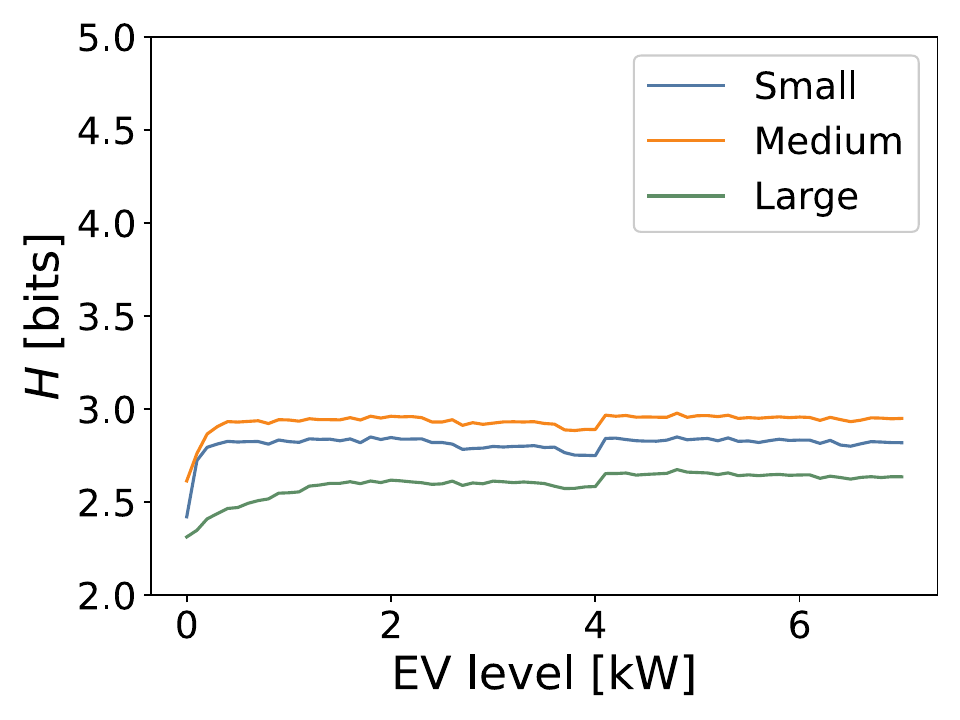}
        % \caption{Mean Shannon entropy for different levels of EV}
        % \label{fig:shannon_ev}
    \end{subfigure}}
    \vspace{-7pt}
    \caption{\small{Trends of Shannon entropy for residential consumers}}
    \label{fig:shannon}
\end{figure}

\begin{figure}[htbp]
    \centering
    \begin{subfigure}[b]{0.96\columnwidth}
        \centering
        \includegraphics[width=\columnwidth]{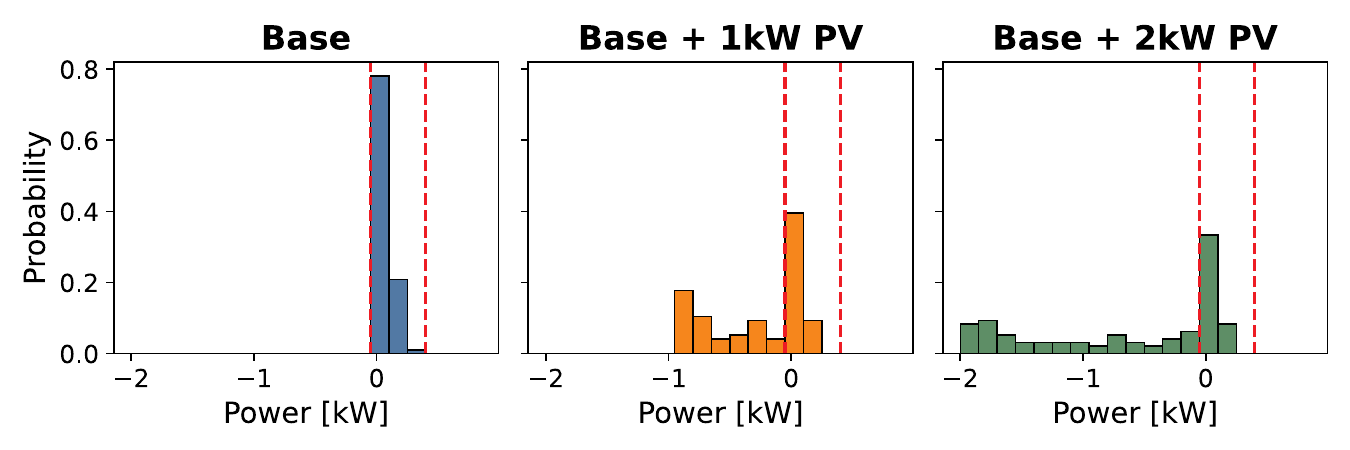}
        % \caption{Different levels of PV penetration}
        % \label{fig:effect PV}
    \end{subfigure}
    \begin{subfigure}[b]{0.96\columnwidth}
        \centering
        \includegraphics[width=\columnwidth]{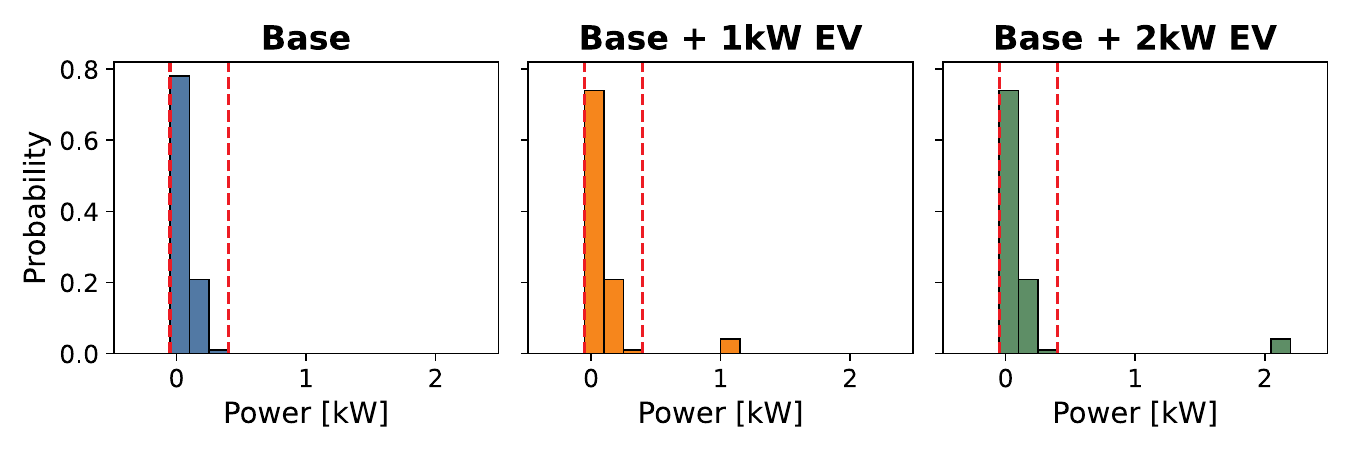}
        % \caption{Different levels of EV penetration}
        % \label{fig:effect EV}
    \end{subfigure}
    \vspace{-7pt}
    \caption{\small{Effect of \textcolor{black}{PV and EV} on the discrete PDF of daily consumption for a small residential consumer. The red dotted lines show the range of base load without EV or PV.}}
    \label{fig:effect_distances}
\end{figure}

The behavior of the Shannon entropy ($H$) is visualized in Fig. \ref{fig:shannon}. 
EV has no significant effect on entropy since it adapts only a minor part of the load profile, resulting in a minimal shift in the distribution of consumption values. PV penetration does impact entropy, but not in direct proportion to the penetration level for the small and medium consumers. For these consumers, entropy increases significantly with the initial additions of PV, then stagnates at higher penetration levels. Fig. \ref{fig:effect_distances} helps explain this behavior.
The early rise in entropy occurs because initial PV integration shifts a large portion of the PDF from high-consumption bins to lower ones. Once daytime consumption is largely offset, further PV additions no longer affect the higher-consumption bins and instead shift the lower-consumption bins more to the left. Since entropy does not take into account the location of the bins, it does not capture these shifts, leading to a stagnation in entropy. This makes the entropy only meaningful at lower levels of PV penetration for the small and medium consumers. Daytime consumption for the large consumer is not offset that fast, making the entropy more meaningful across the entire range of PV penetration levels.

\begin{figure}[htbp]
    \centering
    \begin{subfigure}{0.67\columnwidth}
        \centering
        \includegraphics[width=\columnwidth]{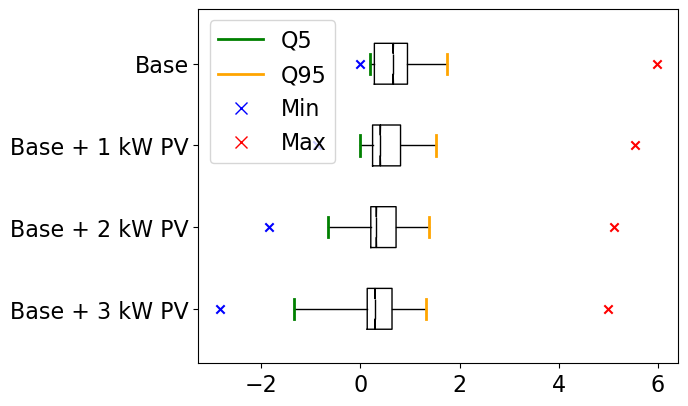}
        \caption{\small{PV integration impact on load}}
    \end{subfigure}
    \begin{subfigure}{0.67\columnwidth}
        \centering
        \includegraphics[width=\columnwidth]{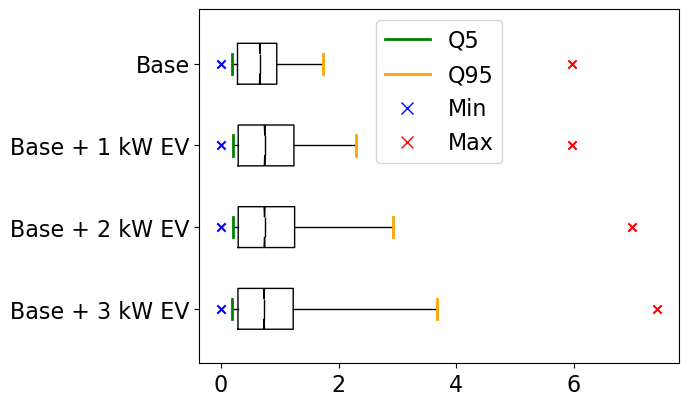}
        \caption{\small{EV integration impact on load}}
    \end{subfigure}
    \caption{\small{Boxplots of yearly profiles for large residential consumers showing minimum, maximum and percentiles}}
    \label{fig:boxplot}
\end{figure}

Fig. \ref{fig:boxplot} shows the effect of \textcolor{black}{PV and EV} on the extreme values in the profile for the large consumer. 
The metrics describing the lower part of the distribution (the minimum, 5th percentile, and lower quartile length) all show a relatively large change with PV and are not affected by EV, as PV reduces a significant portion of the consumption, whereas EV increases only a minor portion.
These metrics are thus useful in quantifying the uncertainty under increased PV penetration.

The metrics describing the upper part of the distribution (the maximum, 95th percentile and upper quartile length) all show a relatively large increase with EV penetration because EV introduces a large peak in the daily profile. 
Since PV reduces a significant portion of the daytime consumption, it also impacts the upper end of the load distribution. However, this effect is mainly limited to the medium and large consumers for the \textcolor{black}{small levels} of PV penetration. For example, metrics like the maximum and 95th percentile decrease initially but stagnate once daytime consumption is fully offset. The left-hand side of Fig. \ref{fig:boxplot} shows this effect for the large consumer.

For small consumers, the daytime consumption is very quickly offset, which limits the effect of PV. In contrast, for medium and large consumers, daytime consumption is not offset as quickly, so PV continues to affect these metrics over a broader range of penetration levels, making them more informative in those cases. On the other hand, the upper quartile length continues to increase even at higher penetration levels. This is because the 75th percentile of daily consumption keeps decreasing with more PV, while the daily maximum remains unchanged once all daytime consumption is offset. As a result, the gap between the 75th percentile and the maximum keeps increasing. These three metrics are thus highly useful under increased EV penetration for all consumers, but only useful under low PV penetration for the medium and large consumers.

The Kullback-Leibler divergence distance ($D_{\text{KLD}}$) is highly sensitive to bins where the probability of the net load is zero while the base load has a non-zero probability, and vice versa. These cases introduce problematic terms in the KLD  because it involve taking the logarithm of the ratio between these probabilities. When either probability is zero, the KLD becomes undefined or extremely large. This occurs frequently for the three sampled consumers. Fig. \ref{fig:effect_distances} illustrates this effect for the small consumer. The region between the two red lines indicates the range of the bins corresponding to the base load PDF. The figure shows that the addition of PV or EV introduces bins outside this range. As this effect is observed across all three consumers, the KLD is ineffective for any of them.

\begin{figure}[htbp] 
    \centering
    \includegraphics[width=\columnwidth]{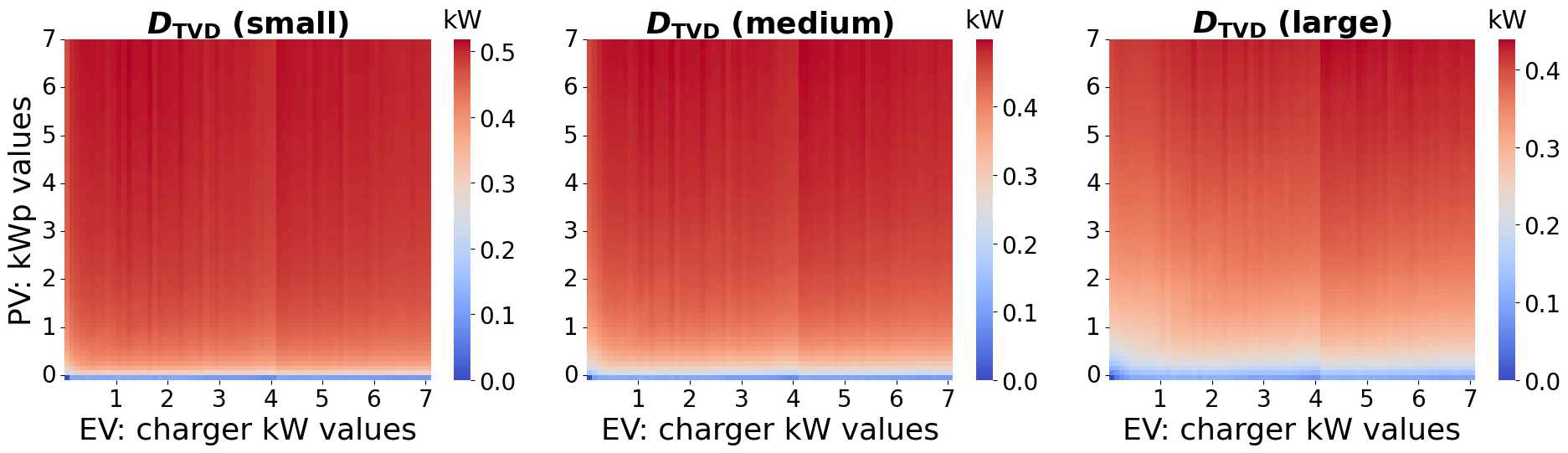}
    \caption{Total variation distance for residential consumers}
    \label{fig:TVD}
\end{figure}

Fig. \ref{fig:TVD} shows the trends for the total variation distance, where the addition of PV initially causes a sharp increase in TVD. However, as the PV penetration level continues to rise, the rate of increase in TVD becomes much lower. 
These trends can be explained by how PV changes the shape of the load profile distribution, as shown in Fig. \ref{fig:effect_distances}. 
When PV is first introduced, it significantly reshapes the distribution, increasing the area (i.e., the TVD) between the PDFs of the base load and net load. As PV penetration increases and daytime consumption is largely compensated, additional PV generation mostly shifts the distribution into lower (more negative) net load values, extending the tail of the distribution. The portion of the distribution that overlaps with the base load (indicated by the red lines) remains largely unaffected. 
Because TVD sums the absolute differences in probabilities, the contribution to the TVD of bins outside the range of the base load comes entirely from the probability of the net load, since the probability of the base load is zero. 

Since the sum of all probabilities of a PDF always sum to one, if the probabilities within the base load range do not change much, then the sum of probabilities outside this range also can not change much. As a result, the TVD between the base load and the net load PDFs shows very minimal changes for increasing PV penetration, \textcolor{black}{see Fig. \ref{fig:TVD}}.
For large consumers, the change in TVD is more gradual, because the overlap between the base and net load distributions lasts longer, making TVD slightly more informative at lower PV levels.
In contrast, EV has little effect on the TVD for all consumers since it affects only a minor part of the distribution. 

\begin{figure}[htbp] 
    \centering
    \includegraphics[width=\columnwidth]{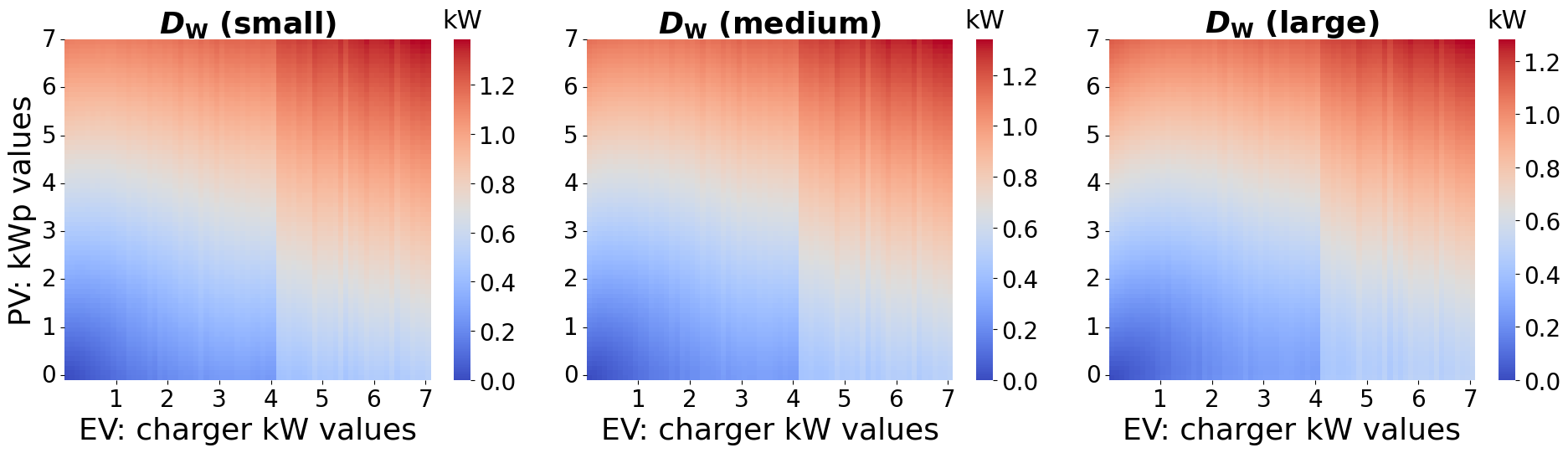}
    \caption{Wasserstein distance between net load and base load}
    \label{fig:wass}
\end{figure}

The Wasserstein distance ($D_W$) captures the full impact of EV and PV on the probability distribution. Fig. \ref{fig:wass} shows the results. On average, each kW of PV shifts about 0.16 kW of probability mass, while each kW of EV shifts around 0.07 kW. These are substantial changes, especially considering the mean daily consumption of 3 residential consumers is less than 1 kW.% making this metric highly useful in quantifying uncertainty under EV and PV penetration for all three consumers.

The error-based metric, MAE, shows a similar absolute increase with both EV and PV as the Wasserstein distance. RMSE also shows the same trend with even higher absolute increases for both, as it places greater weight on larger deviations. 
These metrics, i.e. $D_W$, MAE and RMSE, clearly capture the effects of EV and PV altering the base load profile and are thus highly useful in quantifying the uncertainty for both.

\subsection{Industrial load}
\begin{figure}[htbp]
    \centering
    \includegraphics[width=0.97\columnwidth]{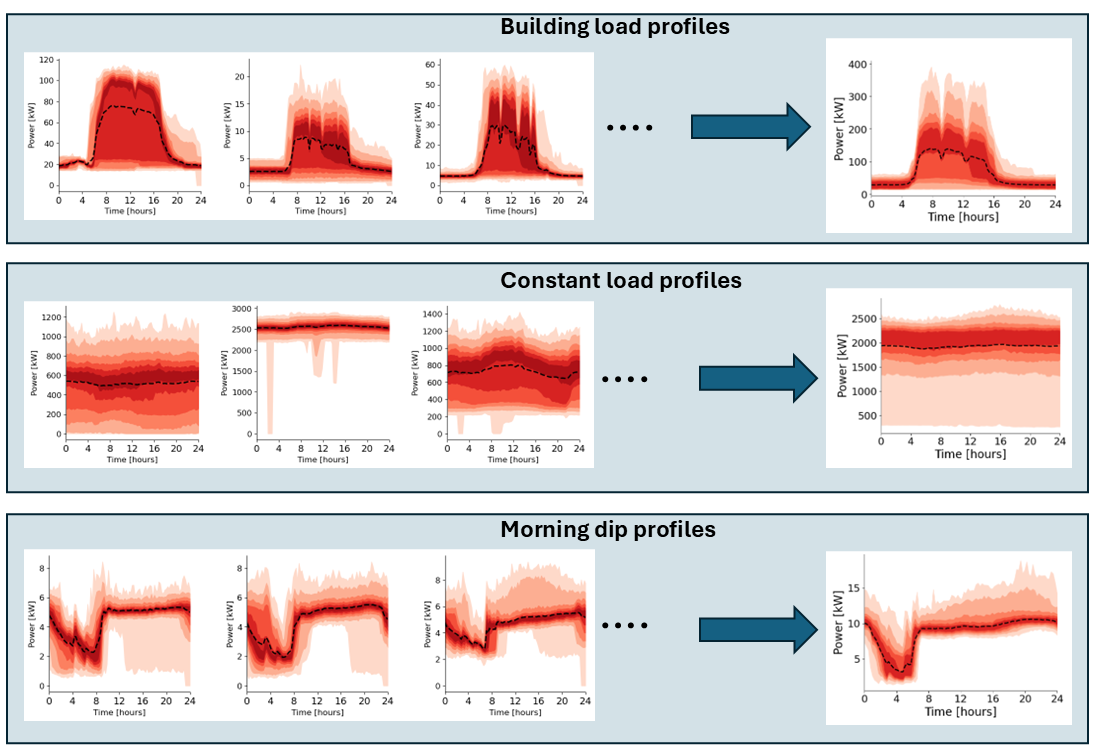}
    \caption{Selecting industrial consumer profiles}
    \label{fig:sel_indus}
\end{figure}

The dataset used for the industrial consumers, contains 50 industrial load profiles with highly diverse consumption patterns and magnitudes \cite{data_industrial}. Three representative consumers are sampled based on their consumption pattern, representing the other profiles following the same kind of consumption pattern. Three largely different consumption patterns are observed. 
Fig. \ref{fig:sel_indus} outlines the procedure we applied for categorizing the industrial load profiles. We used weekday profile distribution for analysis.

\begin{figure}[htbp]
    \centering
    \begin{subfigure}{0.45\columnwidth}
        \centering
        \includegraphics[width=\columnwidth]{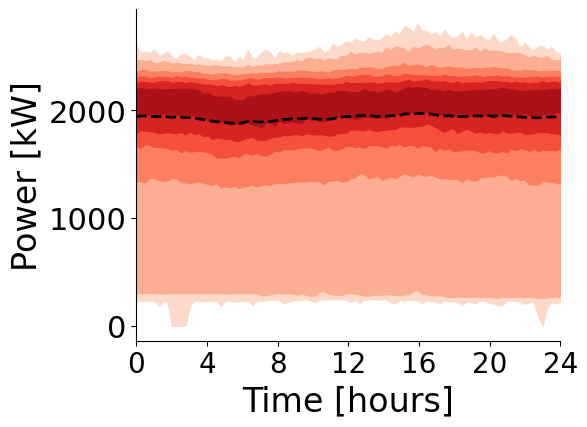}
        \caption{Constant load} \label{fig:constant_load}
    \end{subfigure}
    \begin{subfigure}{0.45\columnwidth}
        \centering
        \includegraphics[width=\columnwidth]{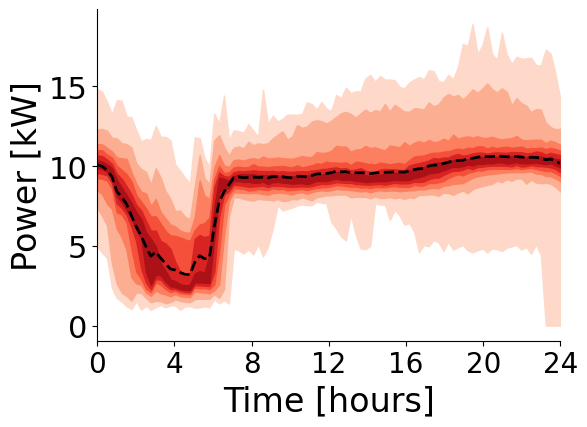}
        \caption{Morning-dip load} \label{fig:valley_load}
    \end{subfigure}
    \begin{subfigure}{0.45\columnwidth}
        \centering
        \includegraphics[width=\columnwidth]{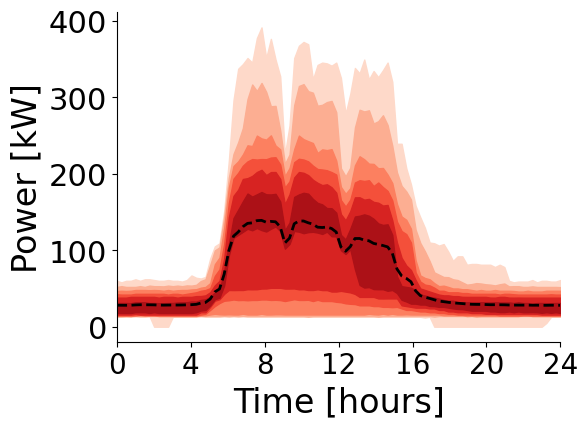}
        \caption{Operational load} \label{fig:building_load}
    \end{subfigure}
    \caption{Representative load profiles of industrial consumers}
    \label{fig:industrial_loads}
\end{figure}

The first type is consumers for which the average daily load stays steady throughout the day and night. They typically have very high overall energy usage.
The second type is profiles, which are characterized by stable daytime and nighttime consumption with a large dip during the early morning hours. They generally represent consumers with lower average power usage.
The third type is profiles, which have a significantly higher consumption during working hours. These profiles are classified as operational loads. The consumers in this group have very varying magnitudes of consumption.
Fig. \ref{fig:constant_load}, \ref{fig:valley_load}, and \ref{fig:building_load} show the sampled profile from the group of high constant consumers, morning-dip consumers, and operational load consumers, respectively.

These profiles are penetrated with PV and EV, increasing linearly from 0 kW to a maximum of 30 kW. Table \ref{tab:metrics_industrial} presents a summary of the calculated metrics for the base case with no DER penetration, the case with maximum PV penetration (30 kW), and the case with maximum EV penetration (30 kW). Type 1, 2, and 3 refer to the constant load, morning-dip load, and operational load, respectively.
\begin{table}[!ht]
    \caption{Result of metrics for industrial consumers}
    \label{tab:metrics_industrial}
    \centering
    \begin{threeparttable}
    \resizebox{\columnwidth}{!}{
    \begin{tabular}{l|ccc|ccc|ccc}
    \toprule
    & \multicolumn{3}{c|}{\textbf{Base}} & \multicolumn{3}{c|}{\textbf{With PV (30 kW)}} & \multicolumn{3}{c}{\textbf{With EV (30 kW)}} \\
    \textbf{Metric} & \textbf{Type 1} & \textbf{Type 2} & \textbf{Type 3} & \textbf{Type 1} & \textbf{Type 2} & \textbf{Type 3} & \textbf{Type 1} & \textbf{Type 2} & \textbf{Type 3} \\
    \midrule
    $C_{\text{annual}}$ & 16,862 & 76 & 584 & 16,820 & 34 & 542 & 16,871 & 84 & 593 \\
    $\sigma$ & 155 & 2.45 & 47.24 & 154 & 6.12 & 46.88 & 155 & 5.35 & 48.65 \\
    $skew$ & -0.61 & -1.39 & 0.70 & -0.62 & -0.58 & 0.33 & -0.61 & 2.51 & 1.62 \\
    $kurt$ & 0.55 & 0.96 & -0.19 & 0.58 & -0.86 & -0.76 & 0.55 & 10.53 & 3.84 \\
    $ramp$ & 42.58 & 0.55 & 6.46 & 42.58 & 0.79 & 6.52 & 42.65 & 1.00 & 6.80 \\
    $H$ & 2.49 & 2.87 & 1.76 & 2.47 & 4.14 & 2.35 & 2.50 & 3.11 & 1.87 \\
    % \addlinespace
    $C_{\text{min}}$ & 0 & 0 & 0 & 0 & -21.18 & -18.40 & 0 & 0 & 0 \\
    $Q_{5}$ & 994 & 2.38 & 14.00 & 983 & -12.81 & 3.00 & 997 & 2.38 & 14.00 \\
    $LQL$ & 255 & 5.90 & 3.05 & 255 & 7.06 & 7.34 & 256 & 5.95 & 3.14 \\
    $C_{\text{max}}$ & 2,798 & 18.90 & 391.00 & 2,774 & 17.97 & 391.00 & 2,798 & 46.16 & 420.90 \\
    $Q_{95}$ & 2,424 & 11.79 & 214.00 & 2,418 & 11.14 & 209.73 & 2,426 & 13.95 & 216.00 \\
    $UQL$ & 121 & 1.46 & 58.98 & 120 & 2.31 & 61.99 & 122 & 21.80 & 67.22 \\
    % \addlinespace
    $D_{\text{KLD}}$ & 0 & 0 & 0 & 0.83 & 3.08 & 1.64 & 0.12 & 0.26 & 0.04 \\
    $D_{\text{TVD}}$  & 0 & 0 & 0 & 0.05 & 0.41 & 0.17 & 0.01 & 0.06 & 0.03 \\
    $D_{\text{W}}$ & 0 & 0 & 0 & 4.84 & 4.84 & 4.84 & 0.97 & 0.97 & 0.97 \\
    MAE & 0 & 0 & 0 & 4.84 & 4.84 & 4.84 & 0.97 & 0.97 & 0.97 \\
    RMSE & 0 & 0 & 0 & 9.02 & 9.02 & 9.02 & 4.93 & 4.93 & 4.93 \\
    \bottomrule
    \end{tabular}}
    \begin{tablenotes}[flushleft]
    \scriptsize
    \item \textit{Units:} $C_{\text{annual}}$ in MWh; $H$ and $D_{\text{KLD}}$ in bits; $D_{\text{TVD}}$, $skew$, and $kurt$ are unitless; others\\ in kW; Type 1, 2 and 3 refers to constant load, morning-dip and operational load profiles.
    \end{tablenotes}
\end{threeparttable}
\end{table} 

The constant industrial consumer shows very small changes in all metrics relative to the base load. This is due to the fact that PV and EV are very small compared to the extremely high baseline consumption. None of the metrics are changed by more than \SI{1}{\percent} at maximal EV or PV penetration relative to the base load. For this reason, most of the metrics are considered not useful for this type of profile. Only the annual consumption is considered useful since it shows some large absolute changes that might still be relevant.

Industrial consumers with a morning-dip load pattern have lower baseline consumption, amplifying the relative impact of both EV and PV integration. In these cases, metrics like annual consumption, standard deviation, and mean ramp show a clear linear relationship with PV penetration and EV charging energy, making them highly effective for quantifying load uncertainty under increased penetration of both DERs.
The operational load also shows notable changes in these metrics, and given the high variability in consumption magnitudes within this consumer group, even greater changes can be expected for other similar consumers since this specific consumer has a relatively high baseline consumption.
The morning-dip profile experiences an increase in standard deviation with higher PV penetration. This is due to the introduction of a second dip around noon, caused by PV generation offsetting daytime consumption, which increases the spread of values across the day. However, for the operational load, PV flattens the profile by offsetting part of the high daytime consumption, resulting in reduced variability and, consequently, a lower standard deviation.

\begin{figure}[htbp]
    \centering
    \includegraphics[width=0.98\columnwidth]{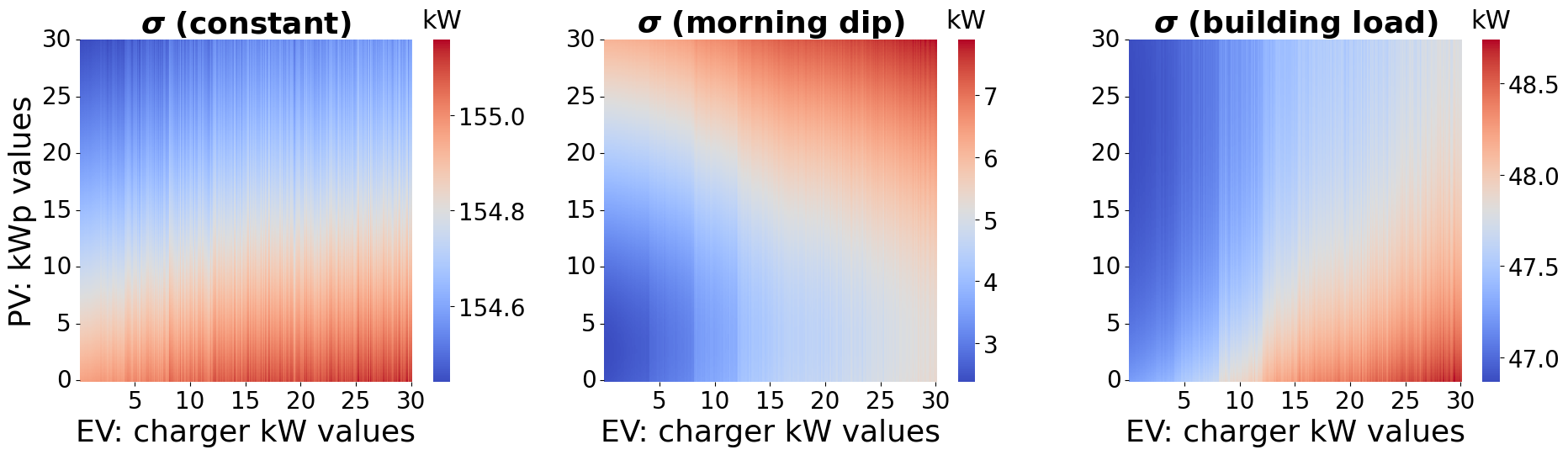}
    \caption{\textcolor{black}{\small{Net load standard deviation for industrial consumers}}}
    \label{fig:industrial_std}
\end{figure}

\textcolor{black}{
The effect of PV penetration on the mean standard deviation varies across the three types of industrial consumers, as shown in Fig. \ref{fig:industrial_std}.
For the constant profile, increasing PV levels lead to a decrease in standard deviation. This is because PV slightly reduces the overall load and smooths out some fluctuations, making the profile more stable. However, the relative effect is very small due to the high baseline consumption.
In contrast, the morning-dip profile experiences an increase in standard deviation with higher PV penetration. This occurs because PV generation significantly offsets midday consumption, creating a second dip in net load around noon in addition to the existing morning dip. As a result, the daily load profile becomes more spread out, increasing its variability.
For the building load, PV flattens the profile by compensating for part of the high daytime consumption, resulting in reduced variability and, consequently, a lower standard deviation.}

% For both consumers, EV charging causes an increase in standard deviation. This is because the added charging load increases a small part of the consumption, leading to greater deviations during those periods.
\begin{figure}[htbp]
    \centering
    \begin{subfigure}[b]{0.49\columnwidth}
        \centering
        \includegraphics[width=1\columnwidth]{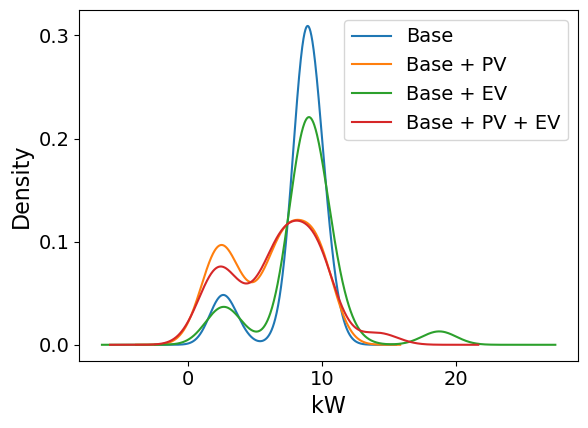}
        \caption{\small{Morning-dip profile}}
        \label{fig:PDF_valley}
    \end{subfigure}
    \begin{subfigure}[b]{0.49\columnwidth}
        \centering
        \includegraphics[width=1\columnwidth]{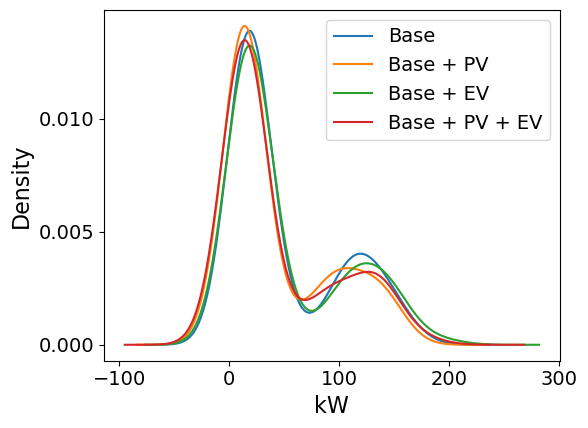}
        \caption{\small{Operational load profile}}
        \label{fig:PDF_building}
    \end{subfigure}
    \caption{\small{Daily PDFs with DER penetration for industrial consumers}}
    \label{fig:PDF_industrial}
\end{figure}

Fig. \ref{fig:PDF_industrial} shows the PDFs of a randomly selected day for the morning-dip and operational load. The PDFs for both consumers have two modes across all scenarios. This limits the interpretability of skewness and kurtosis, as previously discussed for the residential consumers. These two metrics are therefore considered not useful for both consumer types.

\begin{figure}[htbp]
    \centering
    \includegraphics[width=\columnwidth]{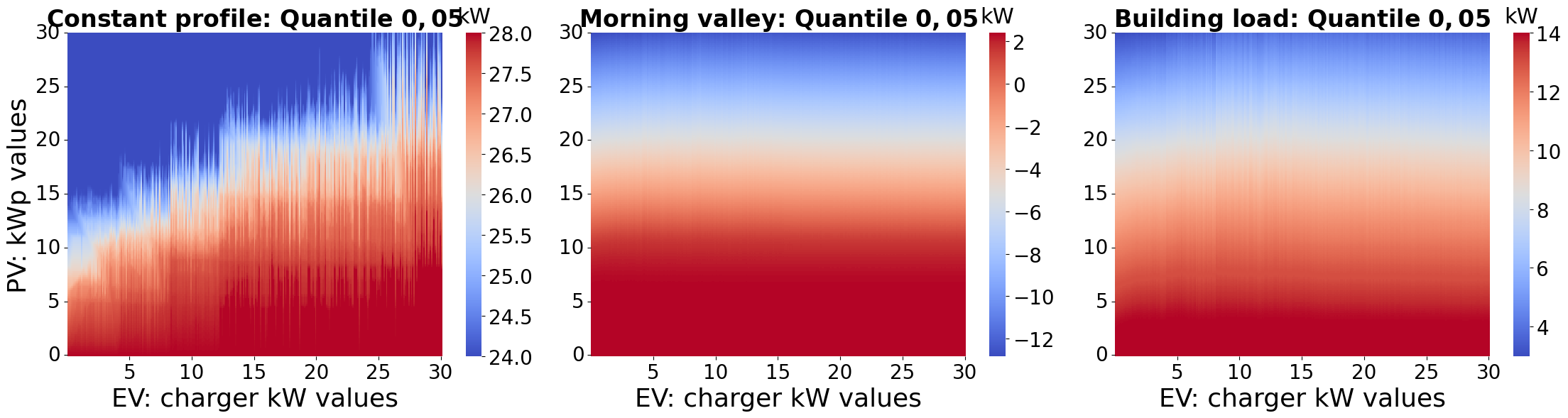}
    \caption{\textcolor{black}{\small{Q5 for industrial consumers}}}
    \label{fig:industrial_q5}
\end{figure}

Metrics that describe the lower part of the distribution, including the minimum, 5th percentile and the lower quartile length, show strong changes with PV for both consumer types, proving their usefulness to quantify uncertainty under increased PV penetration \textcolor{black}{(see Fig. \ref{fig:industrial_q5})}. However, their sensitivity to EV remains negligible, with relative changes of less than  \SI{1}{\percent} compared to the base load at maximal EV penetration for both consumers. 

\begin{figure}[htbp]
    \centering
    \includegraphics[width=0.96\columnwidth]{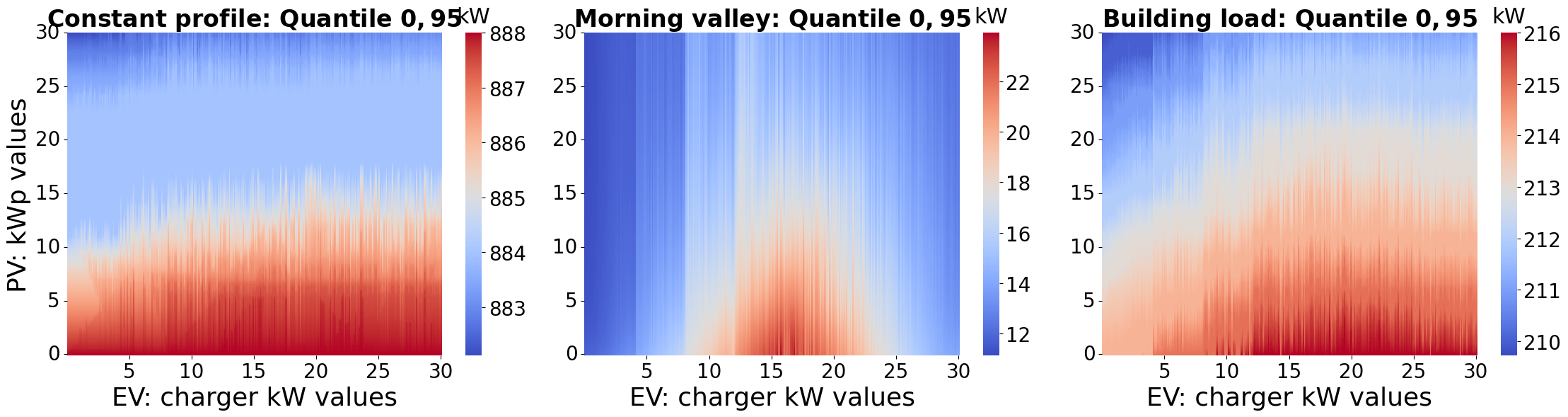}
    \vspace{-7pt}
    \caption{\textcolor{black}{\small{Q95 for industrial consumers}}}
    \label{fig:industrial_q95}
\end{figure}

\begin{figure}[htbp]
    \centering
    \begin{subfigure}{0.6\columnwidth}
        \centering
        \includegraphics[width=\columnwidth]{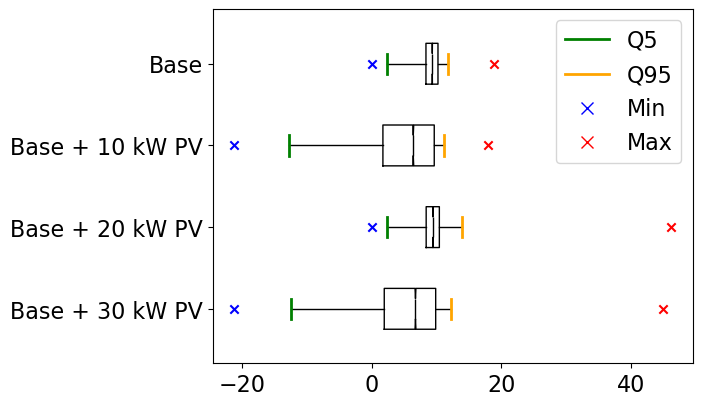}
        \caption{\small{PV integration impact on load}}
    \end{subfigure}
    \begin{subfigure}{0.6\columnwidth}
        \centering
        \includegraphics[width=\columnwidth]{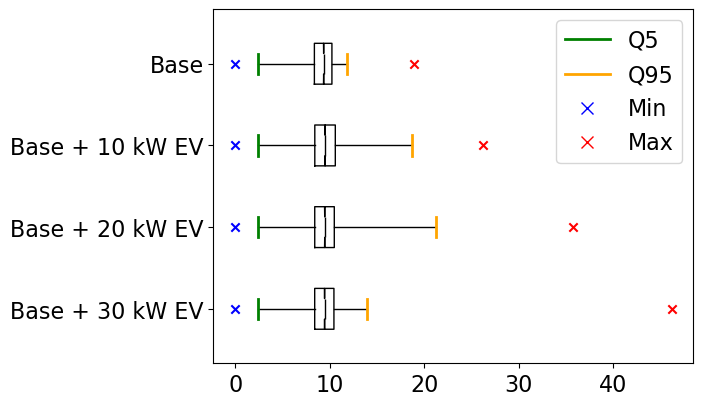}
        \caption{\small{EV integration impact on load}}
    \end{subfigure}
    \caption{\small{Boxplots of yearly profiles for morning-dip industrial consumer}}
    \label{fig:boxplot_valley}
\end{figure}

Metrics characterizing the upper part of the distribution, such as the maximum, 95th percentile, and upper quartile length, show substantial increases in response to EV for both consumers, since it adds a large peak in the profile \textcolor{black}{(see Fig. \ref{fig:industrial_q95})}. These metrics are therefore highly useful under increased EV penetration.
Fig. \ref{fig:boxplot_valley} shows the effect of EV and PV on the extreme values in the profile for the morning-dip consumer. PV clearly impacts the lower end of the distribution, and EV the upper end.

For both consumers, PV also affects the mean daily upper quartile lengths. Although the daily maximum value does not decrease with PV, the 75th percentile does, which spreads the upper quartile and indicates more extreme peaks. This makes this metric also useful under increased PV penetration.

Additionally, for the operational load, PV also affects the 95th percentile, since it reduces a large portion of the daytime consumption. Since the high daytime consumption extends outside solar hours, PV does not decrease the maximum value.

% The influence of PV on these metrics for both consumers is neglectfully small except for the upper quartile length which shows a large relative increase compared to the base loads. This is because daytime consumption is not higher than other consumption values which reduces the impact of PV on the upper part of the consumption. However, PV does lower the 75th percentile, making the higher consumption values appear more extreme. 

\begin{figure}[htbp]
    \centering
    \resizebox{\columnwidth}{!}{
    \begin{subfigure}{0.49\columnwidth}
        \centering
        \includegraphics[width=\columnwidth]{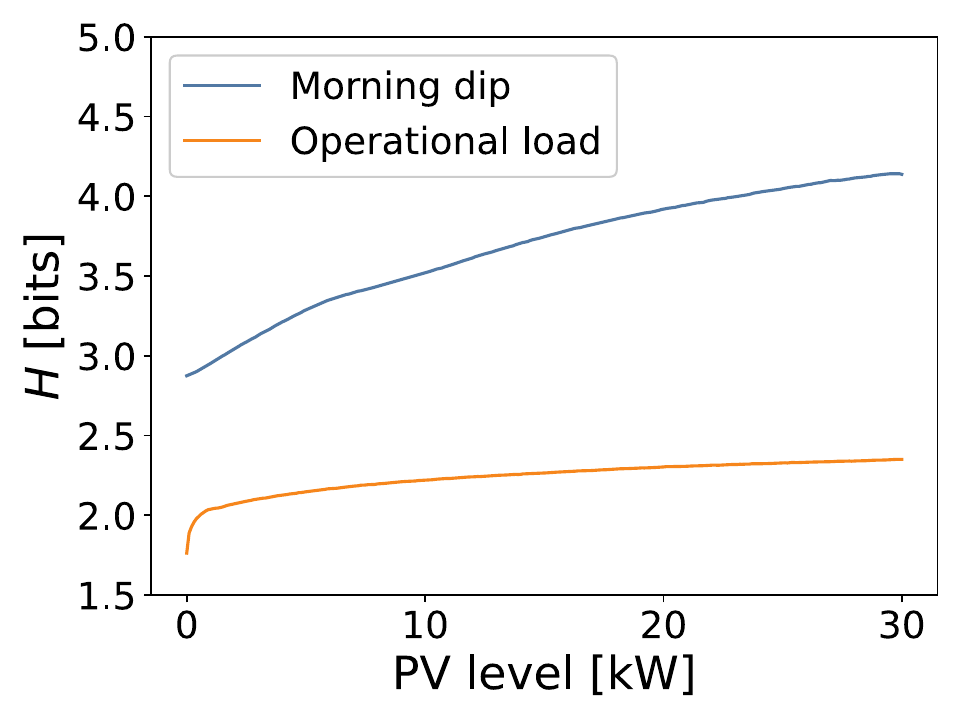}
    \end{subfigure}
    \begin{subfigure}{0.49\columnwidth}
        \centering
        \includegraphics[width=\columnwidth]{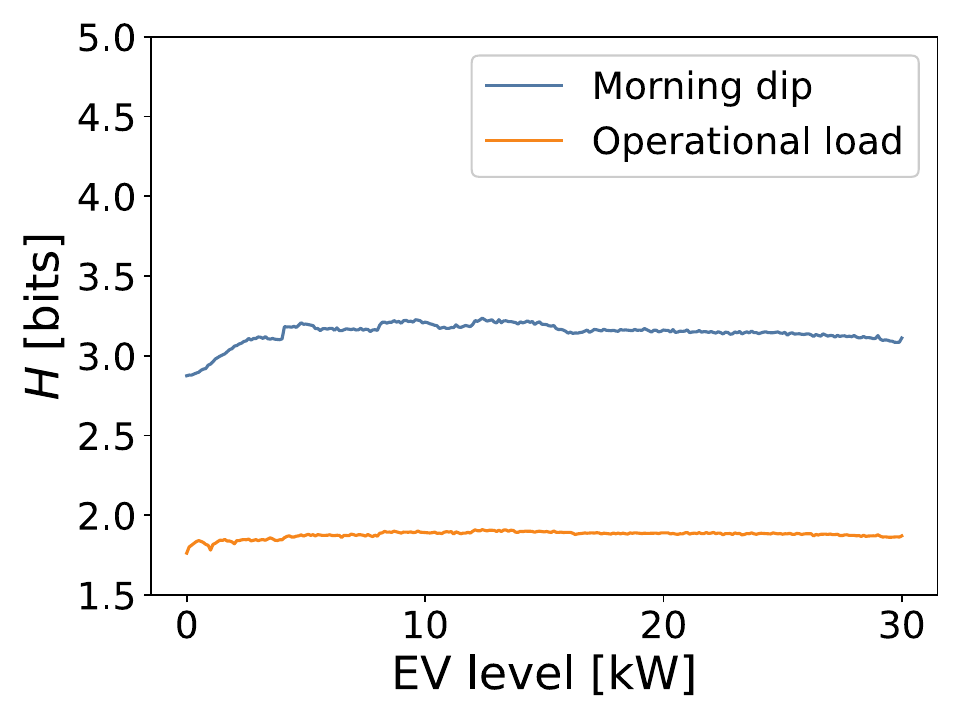}
    \end{subfigure}}
    \caption{Trends of Shannon entropy for industrial consumers}
    \label{fig:shannon_industrial}
\end{figure}

Fig. \ref{fig:shannon_industrial} shows the trends for the mean Shannon entropy with PV and EV. The entropy is, in a proportional manner, increased by PV for the morning-dip profile because the addition of PV continuously redistributes values from high-consumption bins to lower-consumption bins. This makes this metric highly useful in quantifying uncertainty with increased PV penetration.
However, the operational load is not affected by PV in a proportional manner. For these consumers, PV generation primarily reduces consumption during working hours, which results in a shift of the second mode of the net load distribution. Since entropy does not capture shifts in probability bins, it remains largely unaffected and is therefore not considered useful.

The entropy does not show any significant differences for increasing charger kW values for both consumers, making this metric less informative under increased EV penetration.

\begin{figure}[htbp]
    \centering
    \begin{subfigure}[b]{\columnwidth}
        \centering
        \includegraphics[width=0.98\columnwidth]{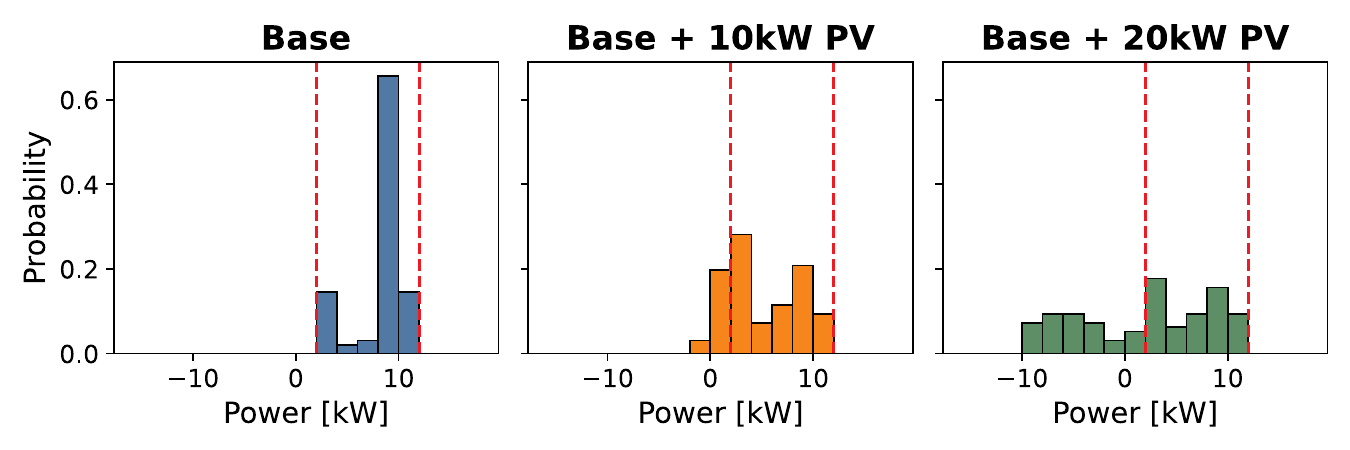}
    \end{subfigure}
    \begin{subfigure}[b]{\columnwidth}
        \centering
        \includegraphics[width=0.98\columnwidth]{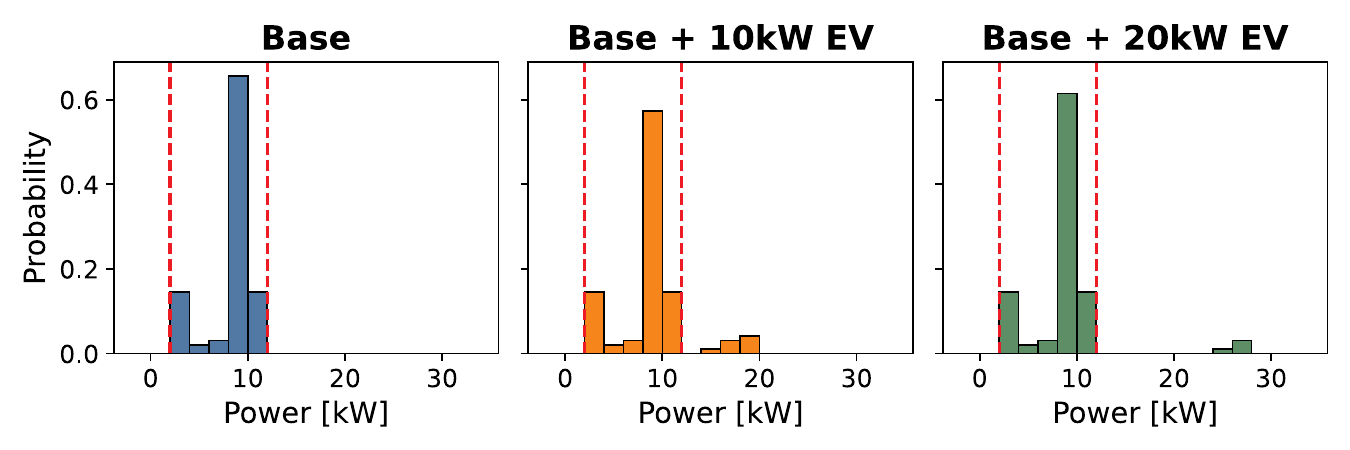}
    \end{subfigure}
    \caption{\small{Effect of \textcolor{black}{PV and EV} on the discrete PDF of daily consumption for the morning-dip profile}}
    \label{fig:hist_valley}
\end{figure}

\textcolor{black}{
For all industrial consumers, the KLD exhibits problematic terms in its calculation for both EV and PV uncertainty.} This is because both DERs introduce bins outside the range of the base load consumption. Fig. \ref{fig:hist_valley} illustrates this effect for the morning-dip consumer. For the operational load, this also frequently occurs, making the KLD a less useful metric for these types of consumers. 

\begin{figure}[htbp] 
    \centering
    \includegraphics[width=0.8972\columnwidth]{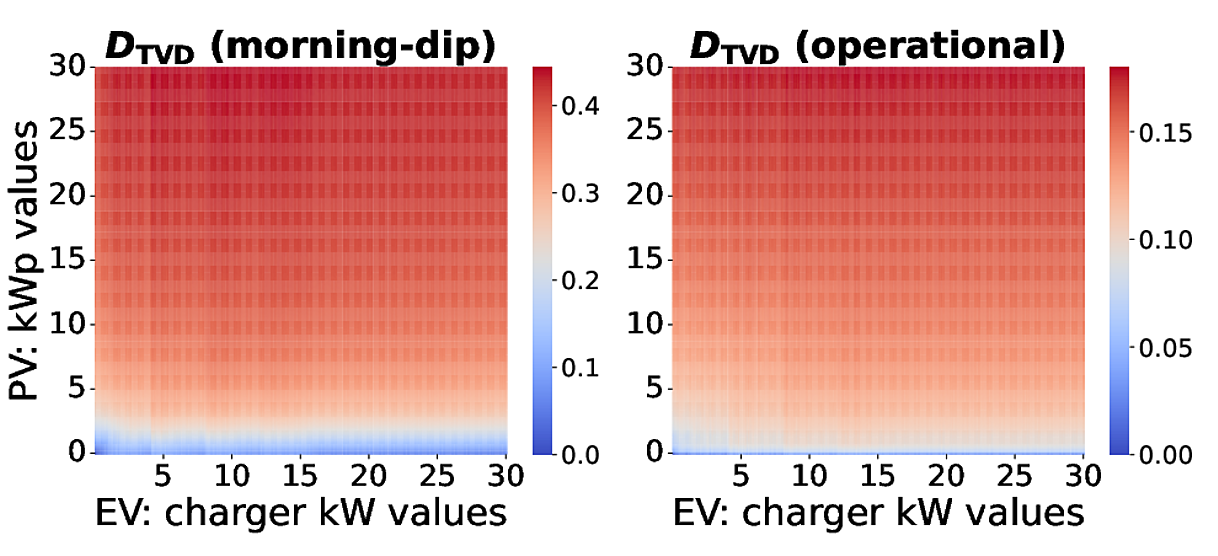}
    \caption{\small{Total variation distance for industrial consumers}}
    \label{fig:TVD_industrial}
\end{figure}

Fig. \ref{fig:TVD_industrial} shows the results for the TVD.
For both consumers, the TVD increases sharply with initial PV integration, as the net load distribution begins to diverge significantly from the base load. However, as PV penetration increases further, the rate of change slows down because the overlapping region of the PDF with the base load remains mostly unchanged. Additional PV primarily shifts the distribution toward more negative consumption values. This effect is illustrated in Fig. \ref{fig:hist_valley} for the morning-dip profile. As a result, the TVD becomes less informative at higher penetration levels. For the operational load, the increase in TVD is less gradual even at low penetration, making this metric less useful overall.
For both profiles, EV shows no significant effect on the TVD since it alters only a very small part of the profile. This is also shown in Fig. \ref{fig:hist_valley}.

\begin{figure}[htbp] 
    \centering
    \includegraphics[width=0.8972\columnwidth]{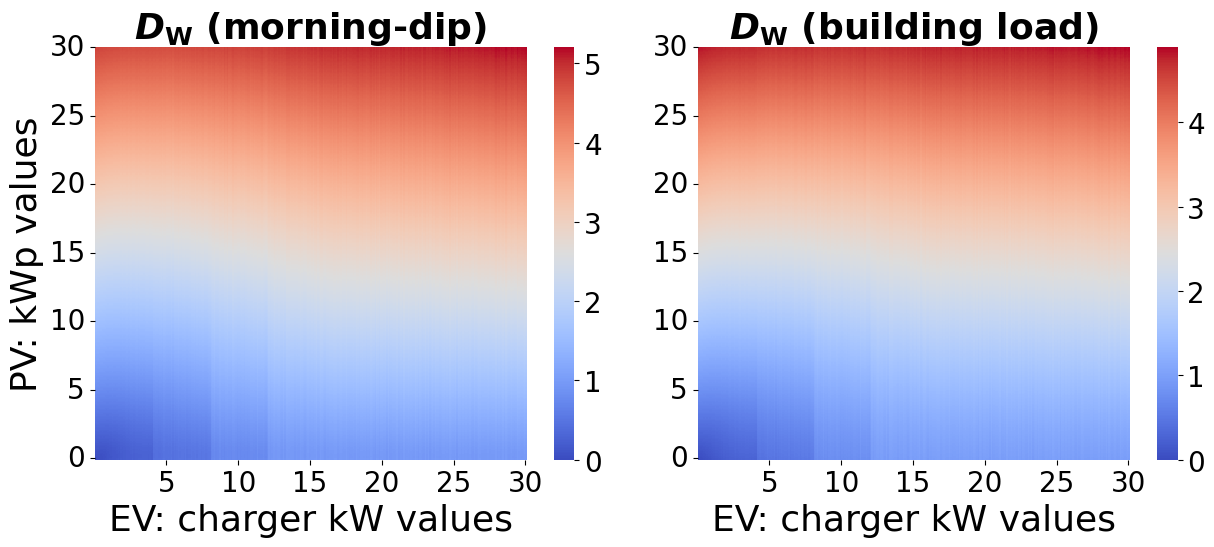}
    \caption{\small{Wasserstein distance for industrial consumers}}
    \label{fig:wass_industrial}
\end{figure}

The Wasserstein distance shows a gradual increase for both PV and EV, effectively capturing the effect of both DERs changing the shape of the PDF for both consumers. Fig. \ref{fig:wass_industrial} shows the results. The effect of PV changing the shape of the PDF is much more pronounced for both, as PV affects a much larger part of the load profile. 
The MAE and RMSE show similar trends with both EV and PV as the Wasserstein distance. This makes the last three metrics useful in quantifying uncertainty for increased penetration of both DERs.

\subsection{Office building load}
\begin{figure*}[htbp]
    \centering
    \begin{subfigure}{0.48\textwidth}
        \centering
        \includegraphics[width=\linewidth]{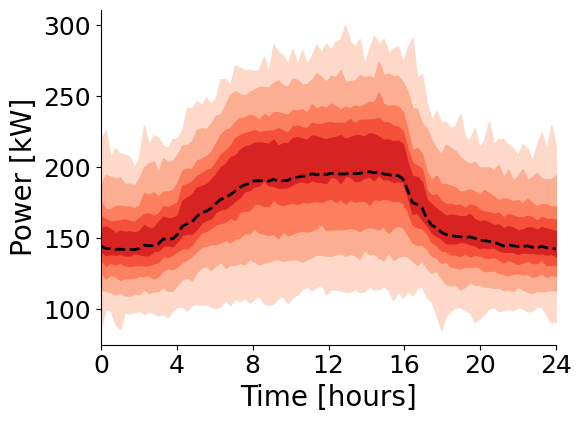}
        \caption{Base load}
        \label{fig:base_comm}
    \end{subfigure}
    \begin{subfigure}{0.48\textwidth}
        \centering
        \includegraphics[width=\linewidth]{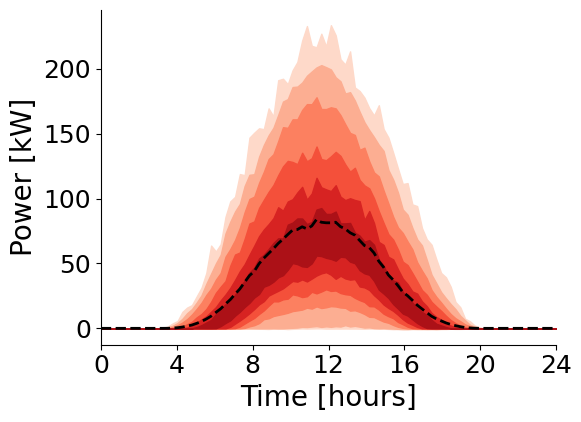}
        \caption{PV generation}
        \label{fig:pv_comm}
    \end{subfigure}
    \begin{subfigure}{0.48\textwidth}
        \centering
        \includegraphics[width=\linewidth]{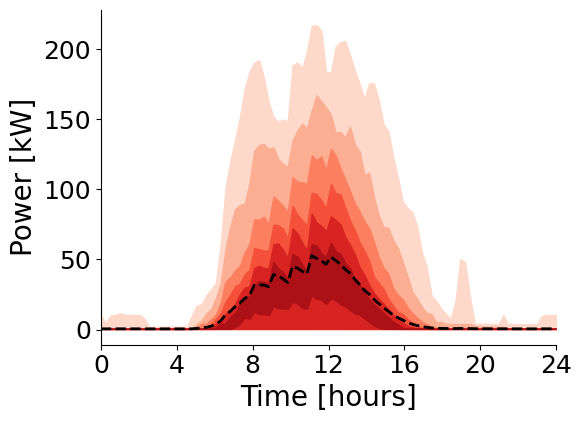}
        \caption{EV consumption}
        \label{fig:ev_comm}
    \end{subfigure}
    \vspace{-7pt}
    \caption{\textcolor{black}{\small{Fancharts of base load (weekday), PV generation and EV consumption of commercial consumer.}}}
    \label{fig:commercial_profiles}
\end{figure*}

For the office building load, data from a single consumer, an office building called EnergyVille, is used. This commercial site has multiple EV charging events per day. Only the aggregated charging profile is available. The available PV generation profile for this site is not normalized to the installed PV capacity. Because this is an office building, there is a clear difference between weekday and weekend power use. On weekdays, the load increases during working hours, while on weekends it stays mostly flat throughout the day. For this reason, only weekdays are included in the analysis. In this analysis, different penetration levels cannot be used due to limited data, which only provides the aggregated EV and non-normalised PV generation profile. The results of the scenarios without DERs, with PV and with EV are shown in Table \ref{tab:metrics_office}. 
\textcolor{black}{
The fan charts for base load, PV generation and EV charging on weekdays are shown in Figures \ref{fig:base_comm}, \ref{fig:pv_comm} and \ref{fig:ev_comm}, respectively.}

\begin{table}[htbp]
    \caption{Results of metrics for office building}
    \label{tab:metrics_office}
    \centering
    \begin{threeparttable}
    \makebox[0.7\columnwidth][c]{%
        \begin{minipage}{0.45\columnwidth} % controls width and aligns notes
        \centering
        \resizebox{\textwidth}{!}{ % textwidth here equals 0.95 * \columnwidth
        \begin{tabular}{lccc}
            \toprule
            \textbf{Metric} & \textbf{Base} & \textbf{With PV} & \textbf{With EV} \\
            \midrule
            $C_{\text{annual}}$ & 1077 & 928 & 1185 \\
            $\sigma$ & 32.21 & 30.15 & 54.27 \\
            $skew$ & 0.10 & -0.22 & 0.52 \\
            $kurt$ & -1.33 & 0.02 & -0.97 \\
            $ramp$ & 10.19 & 13.07 & 11.89 \\
            $H$ & 2.17 & 2.14 & 2.65 \\
            $C_{\text{min}}$ & 0 & -56.50 & 0 \\
            $Q_{5}$ & 115.04 & 82.64 & 115.66 \\
            $LTL$ & 23.53 & 53.88 & 23.62 \\
            $C_{\text{max}}$ & 300.04 & 280.03 & 442.12 \\
            $Q_{95}$ & 242.49 & 213.37 & 316.67 \\
            $UTL$ & 30.08 & 39.11 & 73.43 \\
            $D_{\text{TVD}}$ & 0 & 0.30 & 0.24 \\
            $D_{\text{KLD}}$ & 0 & 20.69 & 0.59 \\
            $D_{\text{W}}$ & 0 & 24.40 & 17.76 \\
            MAE & 0 & 24.59 & 17.89 \\
            RMSE & 0 & 48.22 & 36.87 \\
            \bottomrule
        \end{tabular}}
        \begin{tablenotes}[flushleft]
        \scriptsize
        \item \textit{Units:} $C_{\text{annual}}$ in MWh; $H$ and $D_{\text{KLD}}$ in bits; $D_{\text{TVD}}$, $skew$, and $kurt$ are unitless; others in kW
        \end{tablenotes}
        \end{minipage}
    }
    \end{threeparttable}
\end{table}

\textcolor{black}{
Table \ref{tab:metrics_office} showcases that the annual consumption shows a notable change with EV (10\% increase) and PV (14\% decrease).
The standard deviation decreases by \SI{6}{\percent} with PV and increases by \SI{32}{\percent} with EV.}
Other metrics that show a significant change with both EV and PV are the mean standard deviation and the mean ramp. These changes highlight how EV and PV significantly affect the variability and fluctuations in the office building consumption profile. PV reduces the standard deviation since it offsets some of the high daytime consumption, making the entire profile more flat. 
\textcolor{black}{
EVs the other hand, increase the standard deviation since they make the high daytime consumption even higher.
Both EV and PV contribute to an increase in the mean ramp rate, with EV raising it by \SI{17}{\%} and PV by \SI{28}{\%}. This is because they both have some inherently large ramps which are added to the ramps of the base load.}

\begin{figure}[htbp]
    \centering
    \includegraphics[width=0.75\columnwidth]{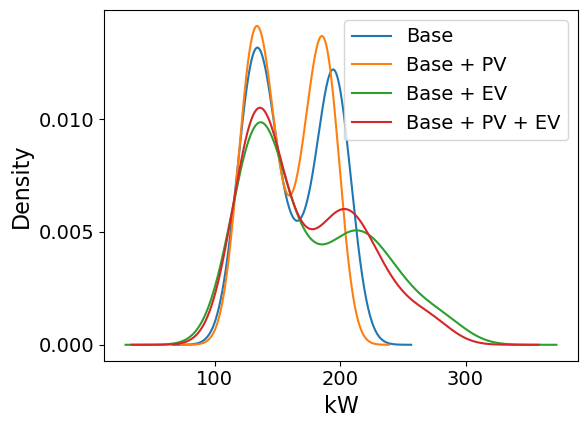}
    \caption{Daily PDFs for office building}
    \label{fig:PDF_office}
\end{figure}

Fig. \ref{fig:PDF_office} shows the daily PDF of a representative day for all four scenarios. All PDFs have two modes, which, once again, limits the interpretation of skewness and kurtosis. This makes them less meaningful for describing the shape of the PDF.

The Shannon entropy is highly increased compared to the base load with EV and decreased with PV. This suggests that adding EVs introduces significantly more randomness or unpredictability into the load profile, flattening or broadening the distribution of values, while PV reduces the randomness or amount of possible consumption values. 
Other consumer types showed stagnating trends with increasing levels of PV penetration for the entropy, and an initial increase and further no differences with increasing levels of EV. Due to limited data availability, the trends can not be investigated for the office building, and thus, no complete interpretation can be made.

However, a possible interpretation is that increasing levels of EV and PV primarily shift the second mode of the distribution. Since entropy does not capture shifts in probability values but rather changes in distribution spread, it is likely not a useful metric for this consumer type.

\begin{figure}[htbp]
    \centering
    \includegraphics[width=0.75\columnwidth]{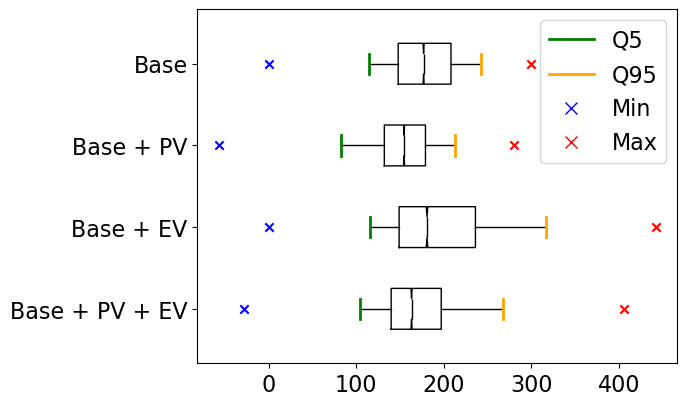}
    \caption{Boxplots of yearly profiles for the office building}
    \label{fig:boxplots_commercial}
\end{figure}

Fig. \ref{fig:boxplots_commercial} illustrates the effect of EV and PV on the extreme values in the profile.
The minimum value and 5th percentile are highly decreased by PV, but remain unaffected by EV. The lower quartile length is significantly increased by PV due to the downward peak introduced, while EV shows no effects. Therefore, these metrics are highly useful with PV penetration, but not with EV penetration.

The metrics characterizing the upper part of the distribution, $C_{\text{max}}$, $Q_{95}$ and $LQL$, are influenced by both EV and PV, where EV increases the high consumption values and PV decreases them. This makes these metrics useful for uncertainty quantification with both EV and PV.

The KLD faces issues in its calculation when either EV or PV is added. When PV is introduced, the second mode of the PDF shifts to the left, as previously illustrated in Fig. \ref{fig:PDF_office}. This leads to bins with zero probability for the net load where the base load does have some probability. The calculation of the KLD for these bins involves taking the logarithm of infinity. When EV is added, this mode shifts to the right, resulting in bins where the net load has a probability but the base load does not. The calculation of the KLD for these bins involves taking the logarithm of zero. Both cases lead to problematic and misleading results. The KLD is therefore not considered useful.

The TVD indicates that both EV and PV change the probability distribution by approximately the same amount compared to the base load. Other consumer types showed a stagnating trend for the TVD with increasing levels of PV. Since the trend cannot be observed due to limited data availability, no conclusive decision can be made about the usefulness.

The Wasserstein distance to the base load is 24 kW with PV and 18 kW with EV, indicating that this amount of power is redistributed across the various bins in the PDF. Since the mean daily power of the base load is $\approx$ 168 kW, these distances represent a significant portion of the PDF that is adjusted. 

The MAE shows a similar absolute deviation from the base load as the Wasserstein distance. The RMSE demonstrates even higher absolute changes for both PV and EV. These last three metrics show substantial effects when EV or PV are added, effectively capturing the shifts in the net load relative to the base load's PDF and load profile.

\subsection{Effect of Sampling Frequency}
\begin{figure}[htbp]
    \centering
    \includegraphics[width=0.9\textwidth]{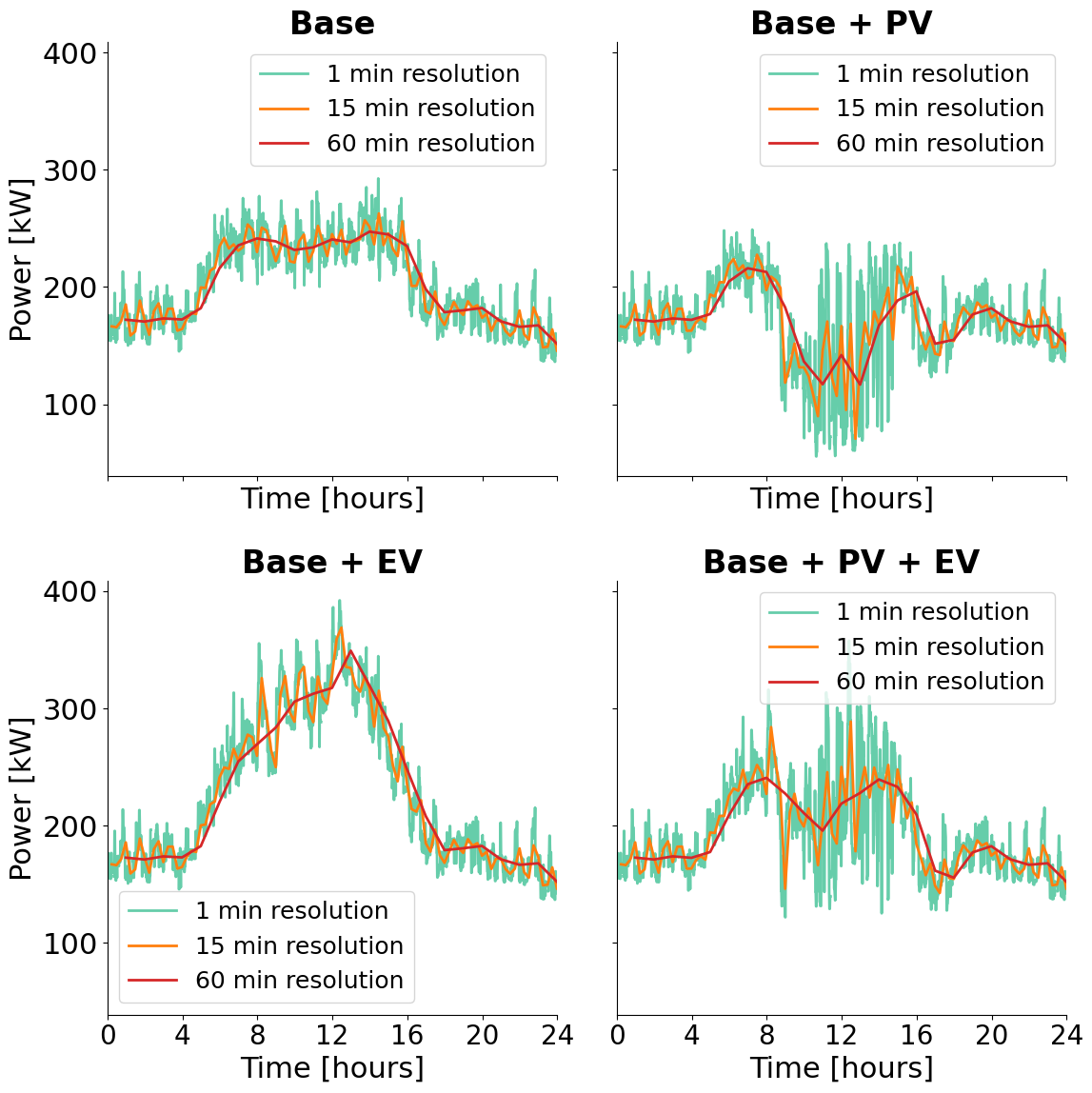}
    \vspace{-7pt}
    \caption{\small{Typical office building load profile for different resolutions}} %'2024-06-4'
    \label{fig:resolutions}
\end{figure}

\textcolor{black}{
This case study for office building load profile examines how the sampling frequency affects the uncertainty in load profiles. This is done by calculating the metrics for the profiles in different resolutions, starting from a resolution in 1-minute intervals and going up to 60-minute intervals. 
To compute the load profile in a specific resolution, the average over the time interval is taken as the consumption value during that time period. Fig. \ref{fig:resolutions} shows the consumption profile of a randomly selected day for each scenario at different resolutions.
As time intervals increase, short-term fluctuations are smoothed out. This averaging effect reduces extreme values. The lower values increase and the higher values decrease, which lowers the peaks and reduces variability.}
\textcolor{black}{
Table \ref{tab:metrics_res} shows the results of the metrics for the profiles in different resolutions, ranging from 1-minute to 60-minute resolution. The results are for the scenario where both EV and PV are present. 
The mean ramp is replaced by the mean ramp rate. The ramp rates are calculated by dividing the mean ramp by the resolution, which provides the ramp rate per minute. This approach allows for a fair comparison of ramps across different time resolutions.}
\begin{table}[htbp]
    \caption{Results of metrics for different resolutions}
    \centering
    \begin{tabular}{lccccc}
    \toprule
        ~ & 1 min & 5 min & 15 min & 30 min & 60 min \\ \midrule
        $C_{\text{min}}$ [kW] & -94.7 & -56.2 & -27.9 & -6.2 & 0 \\ 
        $Q_{5}$ [kW] & 99.4 & 101.9 & 104.4 & 105.5 & 106.6 \\ 
        $LQL$ [kW] & 64.1 & 51.3 & 38.7 & 29.9 & 22.6 \\ 
        \addlinespace
        $C_{\text{max}}$ [kW] & 432.4 & 419.1 & 406.2 & 388.7 & 378.6 \\ 
        $Q_{95}$ [kW] & 271.5 & 269.2 & 267.7 & 266.8 & 263.8 \\ 
        $UQL$ [kW] & 96.2 & 81.2 & 67.9 & 57.6 & 48.2 \\ 
        \addlinespace
        $\sigma$ [kW] & 43.4 & 41.5 & 39.7 & 38.3 & 36.9 \\ 
        \addlinespace
        $ramp\ rate$ [kW/min] & 7.3 & 2.4 & 0.95 & 0.5 & 0.3 \\ 
        \addlinespace
        MAE [kW] & 19.8 & 19.5 & 19.3 & 18.96 & 18.6 \\ 
        RMSE [kW] & 39.4 & 38.4 & 37.5 & 36.7 & 35.9 \\ 
        \addlinespace
        $D_{\text{W}}$ [kW] & 17.62 & 17.69 & 17.73 & 17.72 & 17.65 \\ 
    \bottomrule
    \end{tabular}
    \label{tab:metrics_res}
\end{table}
\textcolor{black}{
Decreasing time resolution drastically impacts the ramp rate
% Increasing the time interval width has the biggest impact on the ramp rates 
with a decrease of \SI{96}{\%} from 1 to 60 minute resolution, see Tab. \ref{tab:metrics_res}. 
% The decrease is the largest for the first few increases in the interval width since this averages out the most extreme fluctuations. After that, the effect of averaging is less drastic since the most significant deviations have already been reduced.
}
\textcolor{black}{
The other metrics follow similar trends, where the largest reductions in uncertainty occur at the first increase in interval width. 
% Following the ramp rate, the next largest changes occur in the minimum value and the mean lower quartile length. 
The minimum increases by \SI{100}{\%}, while the lower quartile length decreases by \SI{65}{\%} from 1 to 60 minute resolution. This indicates that lower values in the profile become less extreme, reducing the impact of downward peaks.}
\textcolor{black}{
The effect of increasing the interval width on the upper part of the distribution is seen most in the UQL, which is decreased by \SI{50}{\%} from 1- to 60-minute resolution, showing that high values are smoothed out and become less extreme compared.}

\textcolor{black}{
In the case of baseline metrics, comparing net load to the base load shows smaller changes of less than \SI{10}{\%} between 1-minute and 60-minute resolutions. This is because averaging affects both the net and base load in a similar manner, preserving most of their relative differences.
% For the other scenarios, base load only, base load with PV, and base load with EV, the overall trends are the same, and the percentage reductions are comparable to those seen in the scenario with both EV and PV. 
% It can be concluded that increasing the time interval significantly reduces uncertainty because larger deviations are smoothed out through averaging. The most substantial reduction occurs with the first few increases in interval width.
}

\pagebreak

% \section{Sensitivity and interaction effects}
\section{Sensitivity analysis}
Sensitivity analysis is a method used to evaluate the impact of uncertainties in input parameters on the outcomes of a model \cite{brevault2020uncertainty}. In this context, the uncertainties are the level of EV and PV penetration, and the outputs are the metrics of the net load. A variance-based sensitivity analysis is performed that identifies which input variables contribute most to the output variability. It decomposes the total output variance into contributions from each uncertain input. The outcome of this sensitivity analysis is a set of sensitivity indices with a value between 0 and 1, where 0 indicates no sensitivity and 1 indicates maximal sensitivity. These sensitivity indices are calculated for each individual consumer in the datasets, and the mean of the indices per group of consumers is given in Table \ref{tab:sens}. %Constant industrial consumers are not considered since the effect of EV and PV on the metrics is very limited.
These results show how EV and PV contribute to load profile uncertainty across different datasets. 
\begin{table}[htbp]
    \caption{Mean sensitivity indices for PV and EV}
    \label{tab:sens}
    \centering
    \resizebox{0.7\columnwidth}{!}{
    \begin{tabular}{{lcccc|cccc}}
        \toprule
        Metric  & \multicolumn{4}{c}{PV sensitivity ($\overline{S_{\mathrm{PV}}}$)} & \multicolumn{4}{c}{EV sensitivity ($\overline{S_{\mathrm{EV}}}$)}  \\ \midrule
        ~ & Res & Ind 1 & Ind 2 & Office & Res & Ind 1 & Ind 2 & Office \\
        $C_{\text{annual}}$ & \textbf{0.77} & \textbf{0.96} & \textbf{0.96} & \textbf{0.65} & 0.19 & 0.04 & 0.04 & 0.35\\ 
        \addlinespace
        $C_{\text{min}}$   & \textbf{1.00}  & \textbf{1.00} & \textbf{0.94} & \textbf{0.81} &  0  & 0  & 0 & 0.09 \\ 
        $Q_{5}$ & \textbf{1.00} & \textbf{1.00} & \textbf{0.94}& \textbf{0.67} & 0  & 0 & 0.01 & 0.18 \\ 
        $LQL$ & \textbf{1.00} & \textbf{0.99} & \textbf{0.96} & \textbf{0.82} & 0 & 0  & 0.03 & 0.09 \\ 
        \addlinespace
        $C_{\text{max}}$ & 0.01 & 0 & 0.08 & 0.04 & \textbf{0.90} & \textbf{0.98} & \textbf{0.68} & \textbf{0.96} \\ 
        $Q_{95}$ & 0.01 & 0.29 & \textbf{0.69} & 0.27  & \textbf{0.94} & \textbf{0.58}   & 0.28 & \textbf{0.72} \\ 
        $UQL$ & 0.02  & 0.02  & 0.08 & 0 & \textbf{0.95} & \textbf{0.97}  & \textbf{0.89} & \textbf{0.98} \\ 
        \addlinespace
        $\sigma$ & 0.36 & \textbf{0.64}  & 0.47 & 0.09  & \textbf{0.65} & 0.34 & \textbf{0.52} & \textbf{0.87}\\ 
        \addlinespace
        $ramp$ & 0.21 & 0.23 & 0.09 & \textbf{0.76} & \textbf{0.65}  & \textbf{0.76} & \textbf{0.90} & 0.23 \\ 
        \addlinespace
        MAE & \textbf{0.81} & \textbf{0.97}  & \textbf{0.97} & \textbf{0.54} & 0.21 & 0.03  & 0.03  & 0.11 \\ 
        RMSE & \textbf{0.54}  & \textbf{0.86} & \textbf{0.86} & \textbf{0.46} & 0.21 & 0.10 & 0.10 & 0.13 \\ 
        \addlinespace
        $D_{\text{W}}$ & \textbf{0.83}  & \textbf{0.97} & \textbf{0.98} & \textbf{0.52}  & 0.20 & 0.03  & 0.01 & 0.07 \\ 
        \bottomrule
        \multicolumn{9}{l}{\footnotesize{Res = residential, Ind 1 = morning-dip industrial, Ind 2 = operational load industrial}} \\
        \multicolumn{9}{l}{\footnotesize{Office = office building (EnergyVille)}}
    \end{tabular}}
\end{table}

% The results show that PV generally has a stronger influence than EV, especially on metrics related to lower consumption levels, due to its consistent daytime impact. EV mainly affects the upper part of the distribution by introducing sharp peaks during charging events. For most consumers, PV dominates the overall uncertainty because of its widespread presence throughout the day, while EV’s impact is more localized. However, commercial consumers show higher EV sensitivity due to frequent daytime charging. Overall, PV alters the entire profile shape more significantly, whereas EV causes sharper, more concentrated variability.

\uline{
The lower part of the distribution, characterized by $C_{\text{min}}$, $Q_5$ and $LQL$, is most sensitive to PV for all consumer types.}
The residential and industrial consumers are almost exclusively sensitive to PV, while the office building shows a slightly lower sensitivity to PV and a higher sensitivity to EV than the other consumers. This can be attributed to the fact that the office building has multiple charging
sessions in one day, increasing all consumption during office hours. When no PV is present, this increased consumption has no effect on the lower part of the distribution since the lowest consumption occurs at night. When PV is present, however, the lowest consumption occurs during the day and adding EV will increase this low consumption. This explains why the office building does show some sensitivity to EV. For the other consumer types, EV charging only happens during really short time periods, limiting the effect of compensating for the decrease in consumption due to PV.

\uline{
The upper part of the distribution, characterized by $C_{\text{max}}$, $Q_{95}$ and $UQL$, is most sensitive to EV for all consumers,} except for the industrial profiles with operational load. For these profiles $Q_{95}$ is highly sensitive to PV. This is because the highest consumption typically occurs during the day for these profiles, when PV generation can directly offset a significant portion of the load. However, the absolute maximum is more sensitive to EV. This is because the increased consumption during the day extends outside solar hours, which results in not all values being decreased by PV, limiting the effect on the absolute maximum. 
The office building also has high daytime consumption, which makes its
sensitivity to PV higher than most other consumer types. However, the effect of EV stays dominant as charging occurs throughout the whole day, increasing a large number of values, unlike other consumers, where charging only happens during a short period of time.

The standard deviation shows a significant sensitivity to both EV and PV for all consumers since they both contribute a lot to the variability in the profile by spreading out consumption values or introducing large deviations.

The mean ramp of the residential and industrial consumers is most affected by EV due to the large charging spikes of the individual charging sessions. PV has inherently smaller ramps in its profile than EV, which makes its effect on the mean ramp lower. The office building, on the other hand, has multiple EVs charging during the day, resulting in an aggregated EV charging profile, which smooths out sharp charging peaks of individual charging sessions. This explains why the mean ramp is more affected by PV in this case.

The overall shape of the load profile and probability distribution (captured by MAE, RMSE and $D_{W}$) is more affected by PV than by EV for all consumers. This is because PV is present for longer periods of time than EV, altering a larger part of the load profile, resulting in a larger overall shift in values compared to the base load.

\pagebreak

\textbf{\textcolor{black}{Sensitivity of base load}}: 
\textcolor{black}{While the previous analysis examined sensitivity at the individual consumer level, where each consumer's base load profile was fixed and only PV and EV levels were varied, this analysis extends the approach by introducing the base load profile as a third uncertain input. This allows analyzing how variability in the selected base load contributes to the overall output variance across a group of consumers. This analysis can not be done for the commercial consumer since only a single base load profile is available.}
\textcolor{black}{In this extended analysis, three uncertain inputs are considered: base load profile, level of PV penetration, and level of EV penetration.
Each of these inputs contributes to the variance in the output metrics. Therefore, the analysis results in three global sensitivity indices: $S_\text{B}$ for the base load, $S_{\text{PV}}$ for PV capacity, and $S_\text{EV}$ for EV capacity.
To compute these indices, multiple simulation iterations are performed in which different base load profiles are sampled and combined with varying levels of PV and EV. This approach quantifies how much each input contributes to the total output variance across the entire consumer group.}
\textcolor{black}{
Unlike the consumer-level analysis, where a separate set of PV and EV sensitivity indices was generated for each consumer (due to the fixed base load), the current analysis returns a single set of sensitivity indices per consumer group. %Since all three inputs are treated as uncertain and varied simultaneously, the resulting indices reflect their global influence across the entire group rather than per individual.
The results are summarized in Table~\ref{tab:sens_incl_cons_all}.}

\begin{table}[htbp]
    \caption{\textcolor{black}{\small{Sensitivity to PV, EV, and base load per consumer type}}}
    \centering
    \resizebox{0.7\columnwidth}{!}{%
    \begin{tabular}{l|ccc|ccc|ccc}
    \toprule
        Metric & \multicolumn{3}{c|}{Residential} & \multicolumn{3}{c|}{Morning-dip} & \multicolumn{3}{c}{Building load} \\
        ~ & $S_\mathrm{PV}$ & $S_\mathrm{EV}$ & $S_\mathrm{B}$ & $S_\mathrm{PV}$ & $S_\mathrm{EV}$ & $S_\mathrm{B}$ & $S_\mathrm{PV}$ & $S_\mathrm{EV}$ & $S_\mathrm{B}$ \\
        \midrule
        $C_{\text{annual}}$ & \textbf{0.64} & 0.16 & 0.18 & 0.06 & 0.00 & \textbf{0.94} & 0.00 & 0.00 & \textbf{1.00} \\
        \addlinespace
        $C_{\min}$  & \textbf{1.00} & 0.00 & 0.00 & \textbf{0.42} & 0.00 & 0.34 & 0.38 & 0.00 & \textbf{0.46} \\ 
        $Q_{5}$ & \textbf{0.99} & 0.00 & 0.01 & 0.26 & 0.00 & \textbf{0.58} & 0.00 & 0.00 & \textbf{1.00} \\ 
        $LQL$ & \textbf{0.99} & 0.01 & 0.01 & 0.06 & 0.00 & \textbf{0.88} & 0.00 & 0.00 & \textbf{1.00} \\ 
        \addlinespace
        $C_{\max}$ & 0.05 & \textbf{0.60} & 0.46 & 0.00 & 0.34 & \textbf{0.64} & 0.00 & 0.00 & \textbf{1.00} \\ 
        $Q_{95}$ & 0.04 & \textbf{0.97} & 0.03 & 0.01 & 0.03 & \textbf{0.96} & 0.00 & 0.00 & \textbf{1.00} \\ 
        $UQL$ & 0.02 & \textbf{0.90} & 0.11 & 0.00 & \textbf{0.96} & 0.02 & 0.00 & 0.00 & \textbf{0.99} \\ 
        \addlinespace
        $\sigma$ & 0.36 & \textbf{0.64} & 0.04 & 0.01 & 0.15 & \textbf{0.84} & 0.00 & 0.00 & \textbf{1.00} \\ 
        $ramp$ & 0.06 & 0.23 & \textbf{0.73} & 0.01 & 0.15 & \textbf{0.84} & 0.00 & 0.00 & \textbf{1.00} \\ 
        \addlinespace
        MAE & \textbf{0.80} & 0.20 & 0.00 & \textbf{0.96} & 0.03 & 0.00 & \textbf{0.96} & 0.03 & 0.00 \\ 
        RMSE & \textbf{0.61} & 0.46 & 0.00 & \textbf{0.86} & 0.10 & 0.00 & \textbf{0.86} & 0.10 & 0.00 \\ 
        $D_{\text{W}}$ & \textbf{0.82} & 0.19 & 0.00 & \textbf{0.98} & 0.02 & 0.00 & \textbf{0.97} & 0.00 & 0.00 \\ 
        \bottomrule
    \end{tabular}}
    \label{tab:sens_incl_cons_all}
\end{table}

\textcolor{black}{The main conclusions from this analysis are that, for residential consumers, uncertainty in the net load is largely driven by EV and PV, as the sensitivity to the base load is low for most metrics. This is because all consumers have a relatively low magnitude of the base load consumption in comparison to the magnitude of the added PV generation or EV consumption. This makes the impact of EV and PV on the profiles very large. Because of this, the uncertainty in the profile will be mainly determined by the level of EV and PV.
The only exception is the mean ramp, which is most sensitive to the base load since there are large differences between different consumers.}

\textcolor{black}{For industrial consumers, the uncertainty in net load is largely determined by the characteristics of the base load profile. Since the base loads in this group are typically large, the relative impact of EV and PV integration is smaller. Differences in the magnitude of consumption between base load profiles lead to significant variation in output metrics, making the choice of base load the most influential factor. This effect is most pronounced for building profiles, which have a very large variety in base load consumption levels. As a result, the uncertainty in the net load is mainly determined by the selected base load profile.
%The metrics with respect to a baseline do not depend on the base load profile. Since the same EV and PV profiles are added, the metrics return the same results for all consumers.
}

\pagebreak

\section{Interaction effects: EV \& PV}
\begin{figure}[htbp]
    \centering
    \begin{subfigure}{0.685\textwidth}
        \centering
        \includegraphics[width=\textwidth]{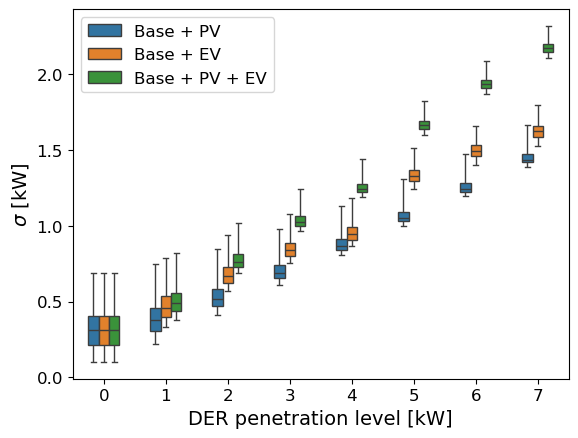}
        \caption{\small{STD distribution for PV, EV and PV+EV case}} %TODO: change title?
        \label{fig:uncertainty_prop_std}
    \end{subfigure}
    \hfill
    \begin{subfigure}{0.686\textwidth}
        \centering
        \includegraphics[width=\textwidth]{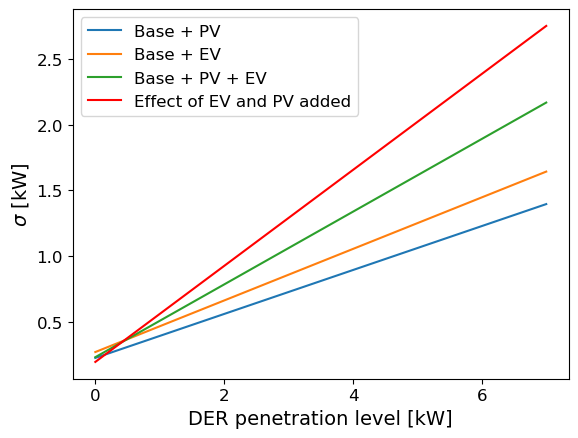}
        \caption{\small{Linear regression through median values}}
        \label{fig:lin_reg_std_res}
    \end{subfigure}
    \caption{\small{Uncertainty propagation per DER for residential consumers}}
    \label{fig:uncertainty_prop}
\end{figure}

This section analyzes how much uncertainty is reduced when EV and PV are considered simultaneously in the net load, rather than adding their effects separately. The percentage change is calculated as follows:
\begin{equation}
\text{Uncertainty reduction} = \frac{z - (x + y - b)}{|x + y - b|} \cdot 100\%,
\end{equation}
% $$\text{reduction} = \frac{z - (x + y - b)}{|x + y - b|} \cdot 100\%$$
\noindent where $b$ is the value of the metric for the base load. $x$ and $y$ are the values of the metric with only PV and only EV penetration, respectively. $z$ is the value of the metric when both EV and PV are considered together in the net load at the same penetration levels as $x$ and $y$.
The reduction at maximal DER penetration levels considered in the analysis is repeatedly calculated for randomly sampled base load profiles and EV charging behaviors. The mean STD of all iterations per consumer group is reported in Fig. \ref{fig:uncertainty_prop} and Table \ref{tab:uncertainty_reduction}.
% Fig. \ref{fig:uncertainty_prop} visualizes this expected result by the red line for the standard deviation.
\textcolor{black}{
The results for the mean standard deviation for the residential consumers are visualized in Fig. \ref{fig:uncertainty_prop_std}, showing the outcome of the simulations across DER penetration levels as boxplots for each case. When both PV and EV are present, the DER penetration level in the \textit{x}-axis refers to the size of PV and EV.
Fig. \ref{fig:lin_reg_std_res} shows the linear regression fit through the median values of standard deviation for the Monte Carlo outcomes. 
The individual effects can be defined as: (i) PV effect: $x - b$, and (ii) EV effect: $y - b$.
% \begin{itemize}
%     \item PV effect: $x - b$
%     \item EV effect: $y - b$
% \end{itemize}
}

\begin{table}[!ht]
    \caption{Percentage change in uncertainty when considering net load instead of adding effects of EV and PV separately}
    \centering
    \resizebox{0.7\columnwidth}{!}{
    \begin{tabular}{lcccc}
    \toprule
        ~ & \makecell{Residential} & \makecell{Industrial \\ Morning-dip} & \makecell{Industrial \\ Operational load} & \makecell{Office \\ EnergyVille}\\ \midrule
        % ~ & small residential & medium residential & large residential & Constant & Moring-dip & Building laodd & Commercial \\ \midrule
        $C_{\text{annual}}$ & 0 & 0 & 0 & 0 \\ 
        \addlinespace
        $C_{\text{min}}$ & 0.09 & 0.01 & 0.84 & 50.66 \\ 
        $Q_{5}$ & 2.56 & 1.70 & 5.28 & 25.33 \\ 
        $LQL$ & 12.17 & 5.43 & 2.70 & 28.28  \\ 
        \addlinespace
        $C_{\text{max}}$ & -0.01 & -0.90 & -0.39 & -3.78  \\ 
        $Q_{95}$ & -30.41 & -8.77 & -0.17 & -6.89  \\ 
        $UQL$ & -0.83 & -22.69 & -7.62 & -17.68  \\ 
        \addlinespace
        \hl{$\sigma$} & -20.94 & -16.57 & -6.07 & -23.95  \\ 
        \addlinespace
        $ramp$ & -0.72 & -0.74 & -0.27 & -3.39  \\ 
        \addlinespace
        $MAE$ & -14.50 & -9.07 & -9.05 & -54.69  \\ 
        $RMSE$ & -36.22 & -30.63 & -30.63 & -55.96  \\ 
        \addlinespace
        $D_{\text{W}}$ & -18.75 & -10.88 & -16.37 & -57.95 \\ 
        \bottomrule
    \end{tabular}}
    \label{tab:uncertainty_reduction}
\end{table}

EV charging tends to partially fill in the valleys introduced by PV production. This raises a part of the low consumption values introduced by PV, as reflected by increases in both $C_{\text{min}}$ and $Q_5$. The office building shows the most pronounced effect, due to EV presence throughout the day, allowing more of the valley of PV to be filled in by EV consumption. This increases more of the low consumption values introduced by PV. The other consumer types have relatively short charging sessions each day, which fill in only a small part of the downward peak introduced by PV, resulting in a less pronounced effect on the lower values.

For residential and industrial consumers, the effect of EV filling in part of the valley of PV results in an increase in the lower quartile length. This happens because the 25th percentile increases, but the daily minimum does not rise as much, since not the entire dip is filled. As a result, the gap between the minimum and the 25th percentile grows, making lower-end deviations more extreme. 
In contrast, for the office building, EV charging tends to align more evenly with PV output throughout the day, reducing the depth of the midday dip more consistently. This leads to fewer extremely low values and consequently a decrease in the LQL, indicating a flatter net load profile with a less pronounced downward peak.

Conversely, PV generation helps offset the peaks caused by EV charging. This reduces metrics such as $C_{\text{max}}$ and $Q_{95}$ and $UQL$. The residential profile in particular shows a larger relative reduction in $Q_{95}$. This is because PV alone has little effect on high consumption values for residential consumers, but when EV is present, PV offsets the peaks introduced by EV. In contrast, for other consumer types, PV already contributes to upper-end reductions even without EV, so the effect of their interaction is small.
The STD ($\sigma$) consistently decreases across all consumer types. This is due to EV and PV cancelling out each other's deviations, resulting in a smoother overall profile. The effect is stronger when EV is present over longer durations, as this allows for offsetting more of the deviations introduced by PV. 
The effects of combining EV and PV on the mean ramps are small. Even when PV offsets part of the energy demand on average, it does not necessarily smooth out the sudden starts or stops of EV charging.

Finally, error-based metrics (MAE, RMSE) and the mean Wasserstein distance show reductions when EV and PV are considered together. Both EV and PV introduce deviations from the base load, but together they compensate for some of each other's deviations, resulting in overall smaller changes when considered together. Once again, the reductions are the largest when EV charging is present for longer durations, aligning more with PV generation.
Overall, the largest reductions are seen when EV consumption can compensate for more of the PV generation. This is the case for the office building where multiple cars are charging, resulting in a continuous charging demand during the day. The EV load thus aligns with PV generation during the day.

\pagebreak

\section{Discussion}
\textcolor{black}{
In this section, we provide a discussion covering 4 topics in the subsequent subsections. Sec. \ref{subsec1} 
% provides the applicability of the UQ metrics, and the data needs have been outlined for different types of metrics.
provides insights into the proposed UQ metrics classification in terms of data needs.
Sec. \ref{subsec2} details how different stakeholders in power networks can utilize the UQ metrics for improved decision making. Sec. \ref{subsec3} summarizes the usefulness of the proposed UQ metrics for different types of load profiles for EV and/or PV growth.
Finally, Sec. \ref{subsec4} conceptualizes an active DN where consumers with DERs can actively reduce uncertainty.
}

\subsection{Proposed UQ classification applicability}
\label{subsec1}
\begin{figure}[htbp]
    \centering
    \includegraphics[width=0.94\columnwidth]{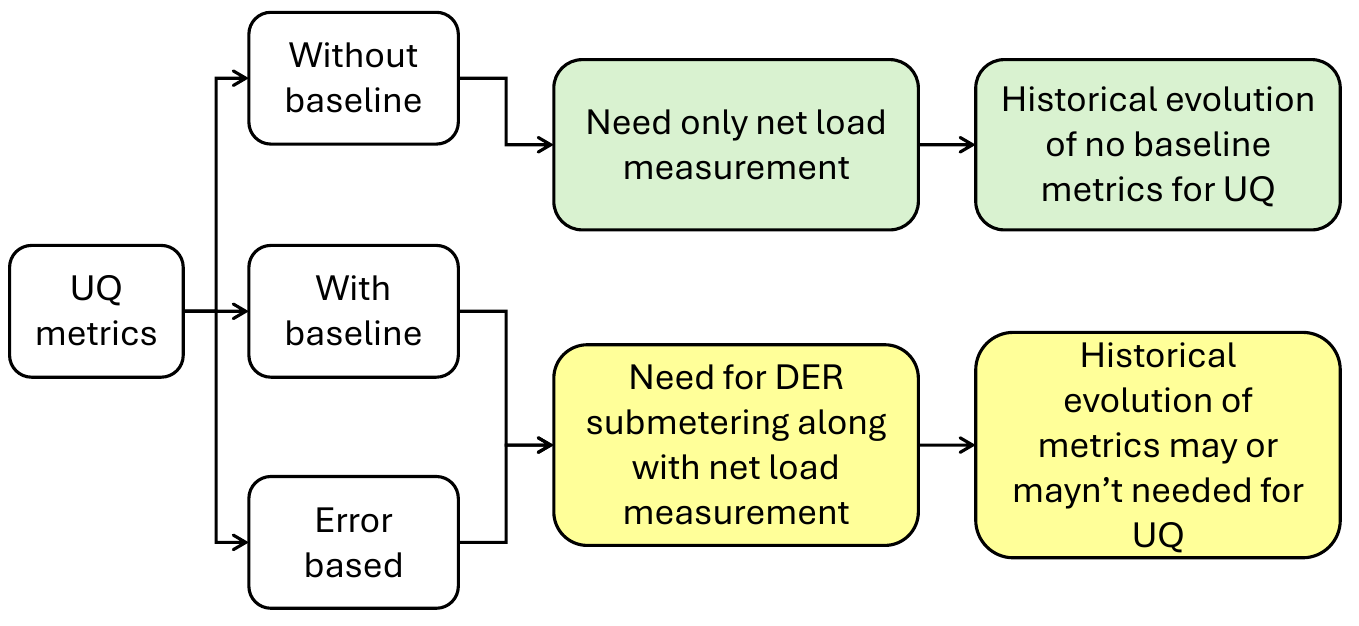}
    \vspace{-6pt}
    \caption{\textcolor{black}{\small{Data needs and interpretability of the proposed UQ metrics}}}
    \label{fig:subsec1_1}
\end{figure}

\textcolor{black}{
UQ metrics evaluate how uncertain consumer load profiles are under different DER penetration levels. With rising EV charging demand and behind-the-meter PV production, load profiles become more stochastic, and UQ helps quantify this uncertainty. The UQ framework is divided into three categories: without baseline, with baseline, and error-based approaches, as proposed in Sec. \ref{sec:metrics}. Refer to Fig. \ref{fig:subsec1_1}.
}

\textcolor{black}{
 \textit{Without baseline approach}: these metrics do not rely on a reference baseline of expected energy consumption. Instead, it uses only the net load measurement, which captures the actual observed load after including both EV load additions and PV generation offsets. This makes the approach practical when baseline consumption patterns are unavailable or hard to estimate. However, it requires examining the historical evolution of no-baseline metrics, meaning past consumption behavior is leveraged to quantify the uncertainty over time. 
}

\textcolor{black}{
\textit{With baseline and error-based approach}: For these metrics, UQ begins with an estimated/measured baseline load profile, i.e., the expected consumption without EVs or PVs. To enhance accuracy, there is often a need for DER submetering (or implementing load disaggregation) along with the net load measurements. Submetering explicitly disaggregates EV charging loads, PV generation, and consumer load demand. Since the baseline exists, these UQ can optionally examine the historical evolution of metrics, which may or may not be necessary depending on how dynamic the system is. For error-based metrics, the focus is on quantifying the error in predictions or forecasts, allowing probabilistic models to capture deviations caused by stochastic DER behaviors.
}

% \textcolor{black}{
% Error-based approach: This method assesses UQ by comparing observed loads with estimates (either with or without a baseline). The focus is on quantifying the error in predictions or forecasts, allowing probabilistic models to capture deviations caused by stochastic DER behaviors. This approach converges with the “with baseline” path in terms of data requirements, including the potential use of submetering, but it is structured around error distributions rather than raw load evolutions.
% }

\subsection{Uncertainty quantification utility for stakeholders}
\label{subsec2}
\begin{figure}[htbp] 
    \centering
    \includegraphics[width=0.95\columnwidth]{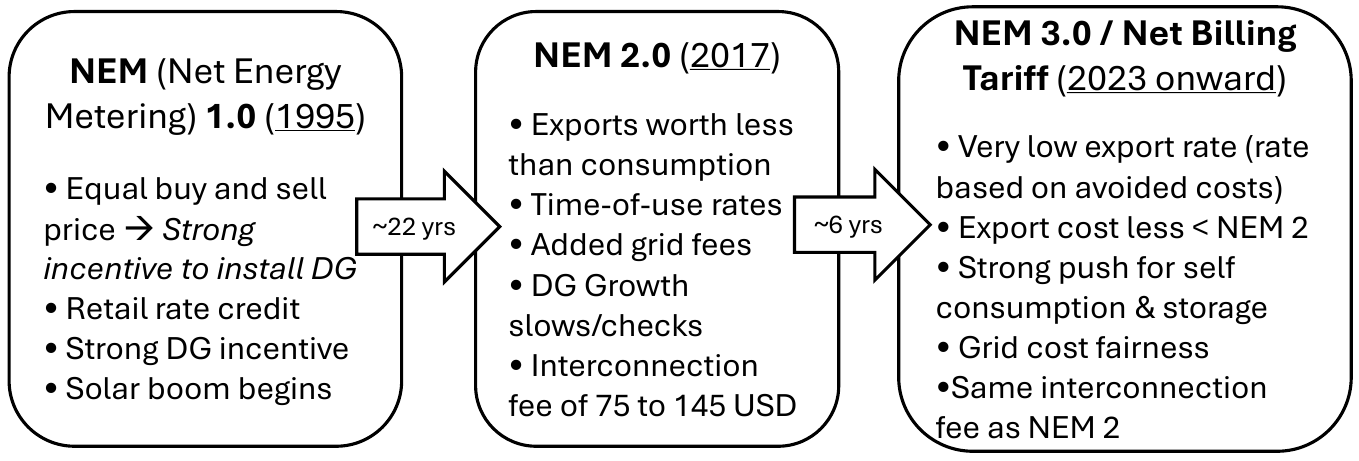}
    \caption{\small{\textcolor{black}{Evolution of net-metering in California drafted by California Public Utilities Commission (CPUC) that is applied for both residential and commercial customers of California's three largest electric utilities—Pacific Gas and Electric (PG\&E), San Diego Gas and Electric (SDG\&E), and Southern California Edison (SCE) \cite{enphaseUnderstandingFuture, cpuc_2022}.}}}
    \label{fig:NEMcalifornia}
\end{figure}

\textcolor{black}{
Next, we detail how accurate UQ can improve decision-making for different stakeholders in DNs.\\
    % \begin{itemize}
$\bullet$ \textit{Distribution system operator}: Using better metrics can assist DSOs in better network planning for future readiness, whether in terms of flexibility planning, network upgrades, or a combination of both. Accurate UQ, can also potentially improve the operational efficiency of DNs by accurate flexibility needs quantification. The second use case of UQ metrics could be for improving DER metadata through better feature selection as performed in \cite{gouveia2024data}.\\
$\bullet$ \textit{Consumers (with/without DERs)}: based on their consumption behaviour and UQ, consumers can select better connection contracts for minimizing their cost of consumption.\\
$\bullet$ \textit{Aggregator/FSP}: the business model of aggregators relies on accurate UQ of distributed aggregated resources. Although the aggregation of a large number of distributed resources becomes fairly predictable as the number of resources aggregated grows, based on the law of large numbers. However, the aggregation of a small number of assets still benefits from better UQ.\\
$\bullet$ \textit{Market operator}: market products incentivizing flexible operation of consumers is expected to grow with growing uncertainty in DNs. Traditionally, peak demand has been penalized in many DNs around the world. However, more sophisticated market products would be needed in future for ensuring reliable operation of the DNs.\\
$\bullet$ \textit{Regulator and policy maker}: Evolution of UQ would govern future connection agreements in DNs. The following example shows how connection contracts evolve in the case of California for new connections of distributed generation (DG):
        As the DG penetration increased, policy adaptations follows, as observed in many parts of the world. See the example of California's Net Energy Metering (NEM) policy in Fig. \ref{fig:NEMcalifornia}.
        This shift reflects California’s attempt to balance DG growth, fairness for non-DG customers, and grid reliability with increased DG penetration.
    % \end{itemize}
}

\pagebreak

\subsection{What metrics to select?}
\label{subsec3}

% \textcolor{red}{In this section, we will summarize the key findings from numerical results and highlight the features of the proposed model.}
This subsection summarizes the key findings from the numerical results. Table \ref{tab:useful_metrics_PV} and \ref{tab:useful_metrics_EV} give an overview of the metrics and their usefulness under increased PV and EV penetration, respectively.
\textcolor{black}{Tab. \ref{tab:useful_metrics_PV} and \ref{tab:useful_metrics_EV}
categorize metrics to "\textit{useful}", "\textit{useful under some scenarios}" and "\textit{not useful}" based on the trend between DER penetration and change in that metric. A highly correlated relationship between a metric under consideration and DER change (EV or PV in the present case) is categorized as "\textit{useful}" and thus a strong indicator. On the other hand, if the metric under consideration does not relatively change with the increase of DER, this is labelled as a "\textit{not useful}" metric. Finally, any other relationship between the metric and DER growth is labelled as "\textit{useful under some scenarios}".}
% Highly useful metrics are indicated in green, non-useful metrics in red and metrics that might be useful under some conditions in yellow.
Constant industrial consumers are clearly an outlier. The base load consumption of these consumers is really high, going up to several GWh per year. This makes the impact of EV and PV, which is in the range of several MWh per year, very minimal. None of the metrics provides any meaningful additional information with increased EV or PV penetration. The constant industrial consumers are not considered in the following discussion when referring to all consumer types.

Among the baseline-free metrics, skewness and kurtosis are not useful for either consumer type due to the bimodal nature of daily profiles. In contrast, the annual consumption, mean standard deviation, and ramp are consistently valuable across all consumer types, reflecting increased variability with rising PV and EV penetration.

Metrics focused on the lower part of the distribution, such as $C_{\text{min}}$, $Q_{5}$, and $LQL$, are particularly useful under PV penetration for all consumer types, capturing the reduction in daytime load. These metrics are less relevant under increased EV penetration. 

Metrics characterizing the upper part of the distribution, $C_{\text{max}}$, $Q_{95}$, and $UQL$, effectively capture peaks introduced by EV for all consumer types, making them useful under increased EV penetration. These metrics can also reflect PV effects for consumers with high daytime loads, such as industrial operational loads and office building, where PV reduces higher consumption levels during the day.

Some metrics, like Shannon entropy, show consumer-specific usefulness. It is only useful for large residential and morning-dip industrial profiles under PV penetration, where it captures the increased spread and randomness introduced by PV. For other consumer types, it is not useful or its usefulness is limited to low PV penetration levels. 

Among baseline-dependent and error-based metrics, Wasserstein distance, MAE, and RMSE are the most effective for all consumer types, reflecting the deviations from the base load profiles, for both PV and EV scenarios. 
KLD and TVD are generally less informative.

Furthermore, when considering both EV and PV together, their combined effect on load profile uncertainty is not additive but often offsetting, leading to significant reductions in uncertainty, particularly for the office building with overlapping EV charging and PV generation. Analyzing the net load is therefore essential for a comprehensive understanding of uncertainty reduction.

\begin{table}[htbp]
    \caption{Overview of useful metrics with PV penetration}
    \vspace{-8pt}
    \label{tab:useful_metrics_PV}
    \centering
    \resizebox{0.8\columnwidth}{!}{
    \begin{tabular}{llllllll}
        ~ & \makecell{Res \\ Small} & \makecell{Res \\ Medium} & \makecell{Res \\ Large} & \makecell{Ind \\ Type 1} & \makecell{Ind \\ Type 2} & \makecell{Ind \\ Type 3} & \makecell{Comm \\}\\\midrule
        $C_{\text{annual}}$ & \cellcolor{green} & \cellcolor{green} & \cellcolor{green} & \cellcolor{green} & \cellcolor{green} & \cellcolor{green} & \cellcolor{green}  \\ 
        \addlinespace
        $\sigma$ & \cellcolor{green} & \cellcolor{green} & \cellcolor{green} & \cellcolor{red} & \cellcolor{green} & \cellcolor{green} & \cellcolor{green} \\
        $skew$ & \cellcolor{red} & \cellcolor{red} & \cellcolor{red} & \cellcolor{red} & \cellcolor{red} & \cellcolor{red} & \cellcolor{red} \\
        $kurt$ & \cellcolor{red} & \cellcolor{red} & \cellcolor{red} & \cellcolor{red} & \cellcolor{red} & \cellcolor{red} & \cellcolor{red} \\
        \addlinespace
        $ramp$ & \cellcolor{green} & \cellcolor{green} & \cellcolor{green} & \cellcolor{red} & \cellcolor{green} & \cellcolor{green} & \cellcolor{green}\\
        \addlinespace
        $H$ & \cellcolor{yellow} & \cellcolor{yellow} & \cellcolor{green} & \cellcolor{red} & \cellcolor{green} & \cellcolor{red} & \cellcolor{yellow} \\
        \addlinespace
        $C_{min}$ & \cellcolor{green} & \cellcolor{green} & \cellcolor{green} & \cellcolor{red} & \cellcolor{green} & \cellcolor{green} & \cellcolor{green}\\
        $Q_{5}$ & \cellcolor{green} & \cellcolor{green} & \cellcolor{green} & \cellcolor{red} & \cellcolor{green} & \cellcolor{green} & \cellcolor{green} \\
        $LQL$ & \cellcolor{green} & \cellcolor{green} & \cellcolor{green} & \cellcolor{red} & \cellcolor{green} & \cellcolor{green} & \cellcolor{green} \\
        \addlinespace
        $C_{max}$ & \cellcolor{red} & \cellcolor{yellow} & \cellcolor{yellow} & \cellcolor{red} & \cellcolor{red} & \cellcolor{red} & \cellcolor{green} \\
        $Q_{95}$ & \cellcolor{red} & \cellcolor{yellow} & \cellcolor{yellow} & \cellcolor{red} & \cellcolor{red} & \cellcolor{green} & \cellcolor{green} \\
        $UQL$ & \cellcolor{green} & \cellcolor{green} & \cellcolor{green} & \cellcolor{red} & \cellcolor{green} & \cellcolor{green} & \cellcolor{green} \\
        \addlinespace
        $D_{\text{KLD}}$ & \cellcolor{red} & \cellcolor{red} & \cellcolor{red} & \cellcolor{red} & \cellcolor{red} & \cellcolor{red} & \cellcolor{red}\\
        $D_{\text{TVD}}$ & \cellcolor{red} & \cellcolor{red} & \cellcolor{red} &  \cellcolor{red} & \cellcolor{yellow} & \cellcolor{red} & \cellcolor{yellow}\\ 
        $D_{\text{W}}$ & \cellcolor{green} & \cellcolor{green} & \cellcolor{green} & \cellcolor{red} & \cellcolor{green} & \cellcolor{green} & \cellcolor{green} \\
        \addlinespace
        $MAE$ & \cellcolor{green} & \cellcolor{green} & \cellcolor{green} & \cellcolor{red} & \cellcolor{green} & \cellcolor{green} & \cellcolor{green} \\
        $RMSE$ & \cellcolor{green} & \cellcolor{green} & \cellcolor{green} & \cellcolor{red} & \cellcolor{green} & \cellcolor{green} & \cellcolor{green} \\
        & \multicolumn{7}{l}{\footnotesize{Green = useful, red = not useful, yellow = useful under some scenarios}} \\
        & \multicolumn{7}{l}{\footnotesize{\textcolor{black}{Type 1: constant, Type 2: morning dip, Type 3: building load}}} \\
    \end{tabular}}
\end{table}
\begin{table}[htbp]
    \caption{Overview of useful metrics with EV penetration}
    \vspace{-8pt}
    \label{tab:useful_metrics_EV}
    \centering
    \resizebox{0.8\columnwidth}{!}{
    \begin{tabular}{llllllll}
        ~ & \makecell{Res \\ Small} & \makecell{Res \\ Medium} & \makecell{Res \\ Large} & \makecell{Ind \\ Type 1} & \makecell{Ind \\ Type 2} & \makecell{Ind \\ Type 3} & \makecell{Comm \\}\\\midrule
        $C_{\text{annual}}$ & \cellcolor{green} & \cellcolor{green} & \cellcolor{green} & \cellcolor{green} & \cellcolor{green} & \cellcolor{green} & \cellcolor{green}  \\ 
        \addlinespace
        $\sigma$ & \cellcolor{green} & \cellcolor{green} & \cellcolor{green} & \cellcolor{red} & \cellcolor{green} & \cellcolor{green} & \cellcolor{green} \\
        $skew$ & \cellcolor{red} & \cellcolor{red} & \cellcolor{red} & \cellcolor{red} & \cellcolor{red} & \cellcolor{red} & \cellcolor{red} \\
        $kurt$ & \cellcolor{red} & \cellcolor{red} & \cellcolor{red} & \cellcolor{red} & \cellcolor{red} & \cellcolor{red} & \cellcolor{red} \\
        \addlinespace
        $ramp$ & \cellcolor{green} & \cellcolor{green} & \cellcolor{green} & \cellcolor{red} & \cellcolor{green} & \cellcolor{green} & \cellcolor{green}\\
        \addlinespace
        $H$ & \cellcolor{red} & \cellcolor{red} & \cellcolor{red} & \cellcolor{red} & \cellcolor{red} & \cellcolor{red} & \cellcolor{yellow} \\
        \addlinespace
        $C_{min}$ & \cellcolor{red} & \cellcolor{red} & \cellcolor{red} & \cellcolor{red} & \cellcolor{red} & \cellcolor{red} & \cellcolor{red}\\
        $Q_{5}$ & \cellcolor{red} & \cellcolor{red} & \cellcolor{red} & \cellcolor{red} & \cellcolor{red} & \cellcolor{red} & \cellcolor{red} \\
        $LQL$ & \cellcolor{red} & \cellcolor{red} & \cellcolor{red} & \cellcolor{red} & \cellcolor{red} & \cellcolor{red} & \cellcolor{red} \\
        \addlinespace
        $C_{max}$ & \cellcolor{green} & \cellcolor{green} & \cellcolor{green} & \cellcolor{red} & \cellcolor{green} & \cellcolor{green} & \cellcolor{green} \\
        $Q_{95}$ & \cellcolor{green} & \cellcolor{green} & \cellcolor{green} & \cellcolor{red} & \cellcolor{green} & \cellcolor{green} & \cellcolor{green} \\
        $UQL$ & \cellcolor{green} & \cellcolor{green} & \cellcolor{green} & \cellcolor{red} & \cellcolor{green} & \cellcolor{green} & \cellcolor{green} \\
        \addlinespace
        $D_{\text{KLD}}$ & \cellcolor{red} & \cellcolor{red} & \cellcolor{red} & \cellcolor{red} & \cellcolor{red} & \cellcolor{red} & \cellcolor{red}\\
        $D_{\text{TVD}}$ & \cellcolor{red} & \cellcolor{red} & \cellcolor{red} &  \cellcolor{red} & \cellcolor{red} & \cellcolor{red} & \cellcolor{yellow}\\ 
        $D_{\text{W}}$ & \cellcolor{green} & \cellcolor{green} & \cellcolor{green} & \cellcolor{red} & \cellcolor{green} & \cellcolor{green} & \cellcolor{green} \\
        \addlinespace
        $MAE$ & \cellcolor{green} & \cellcolor{green} & \cellcolor{green} & \cellcolor{red} & \cellcolor{green} & \cellcolor{green} & \cellcolor{green} \\
        $RMSE$ & \cellcolor{green} & \cellcolor{green} & \cellcolor{green} & \cellcolor{red} & \cellcolor{green} & \cellcolor{green} & \cellcolor{green} \\
        & \multicolumn{7}{l}{\footnotesize{Green = useful, red = not useful, yellow = useful under some scenarios}} \\
    \end{tabular}}
\end{table}

\pagebreak

\subsection{Active reduction in load uncertainty}
\label{subsec4}
\textcolor{black}{
The sensitivity analysis reveals that uncertainty does not scale uniformly with DER penetration. Instead, the dominant source of variability differs by consumer segment: EV adoption contributes more prominently to sharp load ramps and temporal irregularities in residential and industrial loads, whereas PV integration substantially reduces variability in office buildings due to alignment with daytime operational schedules. Importantly, a key finding is that PV and EV, when considered jointly, can partly offset each other’s stochasticity, leading to a reduction in uncertainty. This compensatory interaction \textbf{reduces overall net load uncertainty} and is especially pronounced in office building scenarios, where EV charging demand aligns temporally with PV generation. This observation, although made for an office building, may not scale for other types of consumption load profiles, especially when generation type DERs (PV) and not temporally aligned with load type DERs (EV, heat pumps, etc). However, this alignment can be actively improved by assuming controllable DERs that can be adjusted in time. This can be an outcome of appropriate {congestion market products} \cite{tercca2022deliverable} or {grid availability signal} \cite{hashmi2023perspectives} or {dynamic operating envelopes} (DOEs) \cite{hashmi2023robust2}. Grid availability signals and DOEs can be either centrally, distributed, or decentrally generated.
In some instances, aligning load DERs with DG may not be feasible due to the process and comfort constraints associated with DER operation \cite{getie2025grid}. For such cases, adding energy storage (battery or other storage) can assist in a temporal shift in energy. Thus, leading to an active reduction in load profile uncertainty.
}

\pagebreak

\section{Conclusion}
\label{sec7}
\textcolor{black}{
The rising penetration of distributed energy resources (DERs) introduces growing uncertainty in distribution networks, making it essential to quantify load variability for decisions such as flexibility planning and risk assessment. This study proposes a framework to evaluate net load uncertainty under increasing adoption of photovoltaic (PV) generation and electric vehicle (EV) charging. Using residential, industrial, and office building datasets, the work compares three categories of uncertainty quantification (UQ) metrics: baseline-free, baseline-dependent, and error-based. Findings show that metric suitability depends heavily on consumer type and data availability: baseline-free methods are valuable where reference profiles are unavailable, while baseline- or error-based metrics are preferable when submetering or reliable forecasts exist.
Sensitivity analysis reveals that DER impacts differ across consumer segments. EV adoption drives sharp load ramps and irregularities in residential and industrial contexts, while PV reduces variability in office buildings by coinciding with daytime operations. Notably, joint integration of PV and EV partially offsets individual uncertainties, with the strongest compensatory effect observed in office settings where EV charging overlaps with PV generation.
}

\textcolor{black}{
The study underscores the need to tailor UQ methods to specific consumer behaviors and DER mixes rather than assuming a universal set of UQ metrics. For system operators, aggregators, and other stakeholders, the framework offers practical guidance in selecting UQ tools and interpreting uncertainty dynamics. Incorporating these perspectives into planning and operations can reduce forecasting risks and support resilient, adaptive strategies for evolving DER-rich distribution systems.
}

\pagebreak

\section{Acknowledgments}
This work is supported by the Flemish Government and Flanders Innovation \& Entrepreneurship (VLAIO) through the Flux50 project IMPROcap (HBC 2022.0733), {KU Leuven funded "FlexIQ" project (C2M/24/028)} and  partially funded as DigiRES under CETPartnership-2023 by the European Commission (Grant 101069750) and VLAIO (CETP-2023-00493).

\pagebreak
% This work is supported by the 
% Flemish Government and Flanders Innovation \& Entrepreneurship (VLAIO) through the Flux50 projects InduFlexControl (HBC.2019.0113) and project
% IMPROcap (HBC 2022.0733).
% References
\bibliographystyle{IEEEtran}
\bibliography{reference.bib} % Replace with your .bib file

\end{document}